\documentclass[natbib]{svjour3}

\usepackage{amssymb}
\usepackage{amsmath,amssymb,amsfonts} 
\usepackage{graphicx}
\usepackage{hyperref} 
\usepackage{mathrsfs}
\usepackage{algorithm}
\usepackage{algpseudocode}
\usepackage{algpascal}
\usepackage{geometry}
\usepackage{eurosym}
\usepackage{todonotes}
\usepackage{xcolor}
\usepackage{rotating}

\begin{document}

\title{Data Driven Value-at-Risk Forecasting \\ using a SVR-GARCH-KDE Hybrid\footnote{This is a post-peer-review, pre-copyedit version of an article published in Computational Statistics. The final authenticated version is available online at: \url{http://dx.doi.org/10.1007/s00180-019-00934-7}}}
\author{Marius Lux \and 
            Wolfgang Karl H\"ardle \and
            Stefan Lessmann}

\institute{Marius Lux \at 
            School of Business and Economics\\
            Humboldt-University Berlin\\
            Unter den Linden 6 \\
            D-10099 Berlin, Germany \\
            \email{lux.marius@gmail.com} \and
            Wolfgang Karl H\"ardle \at 
            SKBI School of Business \\
            Singapore Management University\\
            50 Stamford Road, Singapore 178899; \\ 
            C.A.S.E.-Center of Applied Statistics and Economics\\
            Humboldt-University Berlin\\
            Unter den Linden 6 \\
            D-10099 Berlin, Germany \\
            \email{haerdle@hu-berlin.de} \and
            Stefan Lessmann (corresponding author)\at
						School of Business and Economics\\
            Humboldt-University Berlin\\
            Unter den Linden 6 \\
            D-10099 Berlin, Germany \\
						ORCID ID: 0000-0001-7685-262X\\
            \email{stefan.lessmann@hu-berlin.de}}

\maketitle


\begin{abstract}
Appropriate risk management is crucial to ensure the competitiveness of financial institutions and the stability of the economy. One widely used financial risk measure is Value-at-Risk (VaR). VaR estimates based on linear and parametric models can lead to biased results or even underestimation of risk due to time varying volatility, skewness and leptokurtosis of financial return series. The paper proposes a nonlinear and nonparametric framework to forecast VaR that is motivated by overcoming the disadvantages of parametric models with a purely data driven approach. Mean and volatility are modeled via support vector regression~(SVR) where the volatility model is motivated by the standard generalized autoregressive conditional heteroscedasticity (GARCH) formulation. Based on this, VaR is derived by applying kernel density estimation (KDE). This approach allows for flexible tail shapes of the profit and loss distribution, adapts for a wide class of tail events and is able to capture complex structures regarding mean and volatility.
  
The SVR-GARCH-KDE hybrid is compared to standard, exponential and threshold GARCH models coupled with different error distributions. To examine the performance in different markets, one-day-ahead and ten-days-ahead forecasts are produced for different financial indices. Model evaluation using a likelihood ratio based test framework for interval forecasts and a test for superior predictive ability indicates that the SVR-GARCH-KDE hybrid performs competitive to benchmark models and reduces potential losses especially for ten-days-ahead forecasts significantly. Especially models that are coupled with a normal distribution are systematically outperformed. \\
 
\end{abstract}


\section{Introduction}\label{Sec:intro}
Events like the 2008 financial crisis or the outcome of the 2016 referendum in the UK came unexpected for many people. Yet, as these examples illustrate, unlikely events occur at times and they might have far reaching consequences. Risk management is the practice to analyze the macro-environment of an organization, identify possible adverse developments, and design suitable countermeasures.


For financial institutions and systemically important institutions in particular, a key risk management responsibility is to sustain solvency under adverse economic conditions \citep[e.g.,][]{Silva.2017, Kraus.2017}. One of the most popular measures of uncertainty in financial markets is VaR (e.g., \cite{Alexander.2010}). VaR is based on the quantiles of a portfolio's profit and loss ($P\&L$) distribution and can be interpreted as an upper bound on the potential loss that will not be exceeded with a given level of confidence. Its use is appealing because it summarizes the downside risk of an institution in one easily interpretable figure \citep[e.g.,][]{Chen.2012}. 
Regulatory frameworks for the banking and insurance industry such as Basel III or Solvency II also rely on VaR for determining capital requirements. Compared to expected shortfall, an alternative risk measure with some superior mathematical properties \citep[e.g.,][]{Kim.2016}, an advantage of VaR may be seen in the fact that its estimation is more robust due to putting less weight on tail events and large losses, which may deteriorate the quality of statistical estimation routines \citep{Sarykalin.2008}.  


Several approaches have been proposed to estimate VaR including parametric statistical models and data-driven machine learning algorithms such as neural networks (NN) and SVR. In a seminal study, \cite{Kuester.2005} review several statistical methods and compare these in a forecasting benchmark. Using more than 30 years of historical returns data, they find standard GARCH models to forecast VaR with the highest accuracy on average.

GARCH models are also employed by \cite{Chen.2012} to estimate VaR for four daily series of stock market indices. More specifically, \cite{Chen.2012} rely on an asymmetric Laplace distribution and model volatility using a GJR-GARCH model to introduce leverage effects. They then develop a time-varying model to allow for dynamic higher moments. These extensions allow for wider application of the model beyond forecasting. 

Unlike the parametric approach of \cite{Chen.2012}, \cite{Franke.2006} estimate VaR for the German stock index through fitting the mean and volatility of the return series using NNs. More specifically, they model the mean and volatility as an autoregressive (AR) and autoregressive conditionally heteroscedastic (ARCH) process, respectively. To derive VaR and expected shortfall, \cite{Franke.2006} use the predicted mean and variance with the normal distribution. This model outperforms a standard GARCH model in terms of VaR exceedances and proofs capable of quickly adjusting volatility in case of shocks with only short impact. \cite{Dunis.2010} also  propose a NN-based approach towards forecasting VaR and expected shortfall. 

\cite{Khan.2011} develops a VaR-model that forecasts realized volatilities using a combination of a heterogeneous AR model and SVR. VaR is then computed based on the normal, $t$- and skewed $t$-distribution. Applying this model to 5- and 15-minutes return data, \cite{Khan.2011} is able to confirm the suitability of the SVR component.  \cite{Stankovic.2015} provide further evidence that SVR is a useful method for VaR forecasting. Likewise, \cite{Xu.2016} introduce a multi-period VaR model using SVR in a quantile regression framework and show this approach to outperform GARCH models. 

The findings of \cite{Xu.2016} seem to disagree with prior results of \cite{Kuester.2005} where GARCH models predict VaR with highest average accuracy and more accurately than quantile regression approaches in particular. Implementations of the quantile regression using a data-driven SVR model might explain the results of \cite{Xu.2016}. More specifically, the linear and parametric structure of standard GARCH models might be a limiting factor in VaR forecasting. Moreover, the parameters of GARCH-type models are usually estimated via maximum likelihood estimation \citep{Bollerslev.1986}. This necessitates distributional assumptions, which might be problematic since the distribution of financial returns is skewed and exhibits fat tails \citep{Bali.2008,Harvey.2000}. 

Noting the possible limitation of the parametric framework, \cite{Schaumburg.2012} combines extreme value theory with nonparametric VaR estimation to forecast return distributions of four financial stock indices. A parametric conditional autoregressive value at risk (CAViaR) model serves as benchmark. The benchmark and the proposed model both circumvent the estimation of the mean and variance of the $P\&L$ distribution through predicting a quantile directly. In this regard, the approach of \cite{Schaumburg.2012} can be characterized as a nonparametric CAViaR model. 

VaR forecasts based on CAViaR frameworks have also been considered in the benchmarking study of \cite{Kuester.2005}. In fact, the authors also introduce a novel CAViaR model in the paper and test it alongside various other VaR models. However, GARCH models and models relying on the $t$-distribution in particular emerge as most suitable for VaR modeling.


In summary, parametric GARCH models are superior to parametric quantile regression approaches for modeling VaR. To achieve better results with quantile regression approaches, it is necessary to include nonparametric parts into the model. Additionally, data driven GARCH models where the mean and variance components are modeled nonlinearly and nonparametrically lead to better results than parametric GARCH models. However, one shortcoming of the so far proposed data driven GARCH models is the use of parametric residual distributions. Particularly with regard to skewness and kurtosis, this can lead to misspecified residual distributions, resulting in wrong VaR estimates. Therefore, a novel purely nonparametric VaR model that grounds on the GARCH framework is proposed here. We apply data driven approaches to all GARCH components, i.e. mean, variance and residual distribution. Using this approach, we can overcome the possible misspecification of the residual distribution as well as the misspecifications of mean and variance. More specifically, we estimate the mean and variance of the $P\&L$ distribution using SVR and employ KDE to model the density of the standardized residuals  \citep[e.g.,][]{Hardle.2004}. We then integrate these components to derive a VaR forecast. In other words, we propose to start from the most effective parametric modeling approach of \cite{Kuester.2005} and develop models that estimate its components in a purely data-driven manner. In contrast to other so far proposed GARCH based approaches to forecast VaR \citep[e.g.,][]{Youssef.2015, Khosravi.2013, Aloui.2010, Huang.2009, Fan.2008, Chan.2007, Hartz.2006} no assumptions about process dependence structures or the distribution of residuals are made by combining SVR and KDE in a GARCH like fashion. Training the SVR-GARCH-KDE hybrid is, therefore, mainly computationally driven. 

The use of SVR is motivated by the existence of a large body of research showing the effectiveness of SVR in forecasting financial time series \citep[e.g.,][]{Chang.2016, Devi.2015, Tay.2001}. Additionally, \cite{Sheta.2015} compare the forecasting performance of SVR, ANN and traditional linear regression for the S\&P 500. They find that SVR solves the task most successfully. \cite{Kazem.2013} also provide evidence that SVR based models outperform ANNs in the context of financial forecasting. Moreover, \cite{Chen.2009} use a SVR approach for predicting stock market volatility and use a recurrent ANN as benchmark which is outperformed. The use of nonparametric density estimation is motivated by the fact that the existence of fat tails and skewness in the distribution of financial returns can be considered as an empirically proven fact \citep[e.g.,][]{Bali.2008, Harvey.2000}. Although KDE is not a new  approach in VaR forecasting \citep[e.g.,][]{Haerdle.2016, Schaumburg.2012, Malec.2014}, the particular combination of data-driven VaR estimation using SVR and nonparametric density estimation, which we propose in this paper, has, to the best of our knowledge, not been considered in prior work. 

We assess the performance of the proposed model in comparison to GARCH-type models with different error distributions also including skewed and fat-tailed distributions. Empirical experiments using data from three major financial indices, namely the Euro~STOXX~50, Nikkei~225 and Standard \& Poor's 500 (S\&P 500), suggest that the SVR-GARCH-KDE hybrid typically outperforms models that are coupled with a normal distribution and performs competitive to other benchmark models. 

The remainder of the paper is organized as follows. 
In Section \ref{Sec:meths} VaR is defined and the methods underlying the proposed VaR modeling framework are presented. Specifically, the standard GARCH approach, nonparametric density estimation via KDE and SVR are introduced. The proposed SVR-GARCH-KDE hybrid is then developed based on these building blocks. After outlining the theoretical background, the SVR-GARCH-KDE hybrid is compared to other models on different datasets in Section \ref{Sec:emp_study}. Concluding remarks and suggestions for future research are provided in the last section.

\section{Methodology} \label{Sec:meths}



\subsection{Defining Value-at-Risk}

In general, VaR can be derived from the portfolio's $P\&L$ distribution. However, since today's portfolio value is usually known, it suffices to model the return distribution. For a formal description of VaR, let the portfolio returns $r_t$ in period $t$ have the cumulative distribution function (CDF) $F_t$. Then, the VaR in $d$ trading days for a confidence level $1-\alpha$ is defined as

\begin{equation}
VaR_{t+d}^\alpha = -F^{-1}_{t+d}(\alpha) = -\inf\{ x \in \mathbb{R}: F_{t+d}(x) \geq \alpha \} \quad \text{with} \quad \alpha \in (0,1) .
\end{equation}

\noindent In the rest of the paper, VaR refers to the negative $\alpha$-quantile of the next period's portfolio return distribution. 

\subsection{Estimating VaR Using Location-Scale Models} \label{Sec:var_est_approaches}

The proposed VaR modeling framework is based on the location-scale approach. Models of this class estimate the entire distribution of asset returns and derive VaR as a quantile of that distribution \citep[e.g.,][]{Kuester.2005}. Such approaches assume the return process is described as

\begin{equation}
	r_t = \mu_t + u_t = \mu_t + \sigma_t z_t, \qquad z_t \sim (0,1) \ \text{i.i.d.}
	\label{Eq:location_scale_model}
\end{equation}

In (\ref{Eq:location_scale_model}), $\mu_t$ is the location and $\sigma_t > 0$ the scale parameter. Given $r_t$ belongs to the location-scale family and $F_z$ is the CDF of $z$, we can compute VaR as
 
\begin{equation}
	VaR_{t}^{\alpha} = -\left\{\mu_t + \sigma_tF^{-1}_{z}(\alpha)\right\}.
	\label{Eq:location_scale_var}
\end{equation}

\noindent Autoregressive moving average (ARMA) processes and GARCH-type models are commonly used to estimate $\mu_t$ and $\sigma_t$ in (\ref{Eq:location_scale_model}).

\subsection{Modeling Volatility Using GARCH Models}\label{Sec:garch}
\noindent \cite{Bollerslev.1986} introduces GARCH models by generalizing the volatility modeling approach of \cite{Engle.1982}. In deriving the GARCH regression model \cite{Bollerslev.1986} starts by assuming conditional normality of the return process $r_t$:

\begin{equation}
r_t|\mathcal{F}_{t-1} \sim N(\beta^\top x_t, \sigma_t^2).
\label{Eq:arch_cond_norm}
\end{equation}

\noindent where $x_t$ is a vector of lagged endogenous as well as exogenous variables, $\beta$ an unknown parameter vector and $\mathcal{F}_{t-1}$ the information set available at $t-1$. Rewriting (\ref{Eq:arch_cond_norm}) as linear model with conditionally heteroscedastic and normally distributed disturbances gives:

\begin{equation}
r_t = \beta^\top x_t + u_t, \qquad u_t|\mathcal{F}_{t-1} \sim N(0, \sigma_t^2).
\label{Eq:arch_mean_mdl}
\end{equation}

\noindent Then, the GARCH($p$,$q$) representation of the variance $\sigma_t^2$ is
 
\begin{equation}
\sigma_t^2 = \omega + \sum_{i = 1}^{q} \delta_i u^2_{t-i} + \sum_{j = 1}^{p} \theta_j\sigma^2_{t-j}.
\label{Eq:garch_original}
\end{equation}

\noindent \cite{Bollerslev.1986} notes that (\ref{Eq:garch_original}) has an ARMA representation. To see this let $\nu_t = u_t^2 - \sigma_t^2$ and substitute $\sigma_t^2$ in (\ref{Eq:garch_original}) with $u_t^2 - \nu_t$ to obtain

\begin{equation}
u_t^2 - \nu_t = \omega + \sum_{i = 1}^{q} \delta_i u^2_{t-i} + \sum_{j = 1}^{p} \theta_j(u_{t-j}^2 - \nu_{t-j}).
\label{Eq:garch_arma1}
\end{equation}

\noindent Rearranging (\ref{Eq:garch_arma1}) yields an ARMA representation for $u_t^2$:

\begin{align}
u_t^2  &= \omega + \sum_{i = 1}^{q} \delta_i u^2_{t-i} + \sum_{j = 1}^{p} \theta_j(u_{t-j}^2 - \nu_{t-j}) + \nu_t  \\
&=\omega + \sum_{i = 1}^{\max(p,q)} (\delta_i + \theta_i) u_{t-i}^2 - \sum_{j = 1}^{p} \theta_j \nu_{t-j} + \nu_t. 
\label{Eq:garch_arma2}
\end{align}

\noindent Based on (\ref{Eq:garch_arma2}) nonlinear and nonparametric volatility models can be introduced. This can be seen by noting that the conditional expectation of $u_t^2$ is equal to $\sigma^2$. Consequently, the variance process can be modeled solely based on the observed values without making assumptions about the distributional form of the residuals or the structure of the variance process. Hence, the volatility model in the SVR-GARCH-KDE hybrid is motivated by the ARMA representation of $\sigma^2$.

\subsection{Nonparametric Density Estimation}\label{Sec:NDE}

 The volatility of stock returns varies over time and a similar behavior has been observed for the third and fourth moment of the return distribution. For example, \cite{Bali.2008} show that VaR forecasts can be improved by using past estimates of skewness and kurtosis. Given the evidence for the leptokurtic nature of stock returns \citep{Franke.2011}, parametric distributional models might lack the flexibility to capture such distributional characteristics, which motivates the use of nonparametric methods such as KDE  \citep[e.g.,][]{Hardle.2004}.
 


Let $X$ be a random variable with an absolutely continuous distribution function $F$. Further, denote the corresponding density function as $f$ and let $\{x_1, \dots, x_n\}$ be a sample of i.i.d. realizations of $X$. Then, the kernel density estimator $\hat{f}_h(x)$ of $f(x)$ is defined as

\begin{equation}
\hat{f}_h(x) = \dfrac{1}{hn} \sum_{i = 1}^{n} K\left(\dfrac{x_i - x}{h}\right) 
\label{Eq:kde_definition} 
\end{equation}

\noindent where $h$ is a bandwidth parameter with $h>0$ and $K$ is a so-called kernel function. 
Usually, a kernel function is assumed to be a symmetric density function, i.e.  

%
%


\begin{equation}
	\int_{-\infty}^{\infty} K(u)du = 1 \quad \text{with} \ K(u) \ge 0
	\label{Eq:kernel_function_density}
\end{equation}

\noindent and

\begin{equation}
\int_{-\infty}^{\infty} uK(u)du = 0.
\end{equation}

\noindent Conveniently, (\ref{Eq:kernel_function_density}) implies that $\hat{f}_h(x)$ is also a density. Note that $\hat{f}_h(x)$ inherits all properties of $K$ regarding continuity and differentiability.

The KDE based quantile estimator to forecast VaR can be derived as follows. First, the  estimator for $F(x)$ that is based on KDE needs to be derived. Denote $\widehat{F}_h(x)$ as the KDE based estimate of $F(x)$. Then, $\widehat{F}_h(x)$ can be derived as follows:

\begin{align}
\widehat{F}_h(x) &= \int_{-\infty}^{x} \hat{f}_h(z)dz \\
&= \int_{-\infty}^{x} \dfrac{1}{nh} \sum_{i = 1}^{n} K\left(\dfrac{z - x_i}{h}\right)dz \\
&= \dfrac{1}{nh} \sum_{i = 1}^{n} \int_{-\infty}^{x}  K\left(\dfrac{z - x_i}{h}\right)dz.
\end{align} 

\noindent Since the given kernel function $K$ is a density, let $\Gamma$ denote the corresponding CDF. Moreover, using the substitution $u = (z - x_i)/h$ one obtains

\begin{align}
\widehat{F}_h(x) &= \dfrac{1}{n} \sum_{i = 1}^{n} \int_{-\infty}^{\frac{x - x_i}{h}}  K\left(u\right)du \\ 
 &= \dfrac{1}{n} \sum_{i = 1}^{n} \Gamma \left( \dfrac{x - x_i}{h} \right).
 \label{Eq:kde_cdf_est}
\end{align} 

\noindent Thus, $\widehat{F}_h(x)$ is the mean of the CDF corresponding to $K$ evaluated at $(x-x_i)/h$ for $i = 1, \dots, n$. Then, for $\alpha  \in (0,1)$ the KDE based quantile function $\widehat{Q}_h$ is obtained as

\begin{equation}
	\widehat{Q}_h(\alpha) = \widehat{F}_h^{-1}(x).
\end{equation}

\subsection{Support Vector Regression} \label{Sec:meths_svr}
SVR can be understood as a learning method to solve nonlinear regression tasks  \cite[e.g.,][]{Smola.2004}. It shares some similarities with a three-layer feed-forward NN and is able to approximate arbitrarily complex functions \citep{Chen.2009}. However, NNs are based on minimizing the so-called empirical risk and tend to find only locally optimal solutions. In contrast, SVR minimizes the so-called structural risk to achieve better generalization and solves a convex optimization problem leading to a globally optimal solution.
To describe the SVR model, let $\{(y_i, x_i)| i = 1,\dots, n; \ n\in \mathbb{N}\}$ with $x_i \in \mathbb{R}^p$ and $y_i \in \mathbb{R}$ denote the training data set. Suppose $f$ is a linear function such that 

\begin{equation}
f(x) = \omega^\top x + b
\label{Eq:svr_lin_base_mdl}
\end{equation}

\noindent where  $\omega \in \mathbb{R}^p$ and $b \in \mathbb{R}$. Then, SVR aims to find an approximation of $f$ that deviates at most by $\epsilon$ from the observed target $y$ while being as flat as possible (i.e., in the sense that weights in $\omega$ are small). This translates into the following convex optimization problem:

\begin{equation}
\begin{aligned}
& \text{minimize} & & \frac{1}{2} \|\omega\|^2 \\
& \text{subject to} & & \begin{cases}
y_i - \omega^\top x_i - b \le \epsilon  \\
\omega^\top x_i + b - y_i \le \epsilon.
\end{cases} 
\end{aligned}
\label{svm_mean_eq}
\end{equation}
\newline
In view that (\ref{svm_mean_eq}) might lack a feasible solution, \cite{Vapnik.1995} introduces an $\epsilon$-insensitive loss function: 

\begin{equation}
L_{\epsilon}\left\{y - f(x)\right\}  = \begin{cases}
0 \qquad \qquad \qquad  \text{if} \ |y - f(x)| \le \epsilon \\
|y - f(x)| - \epsilon \ \ \text{otherwise}.
\end{cases}
\label{Eq:svr_alt_loss_fct}
\end{equation}

\begin{figure}[h!]
	\centering
	\includegraphics[scale = 0.4]{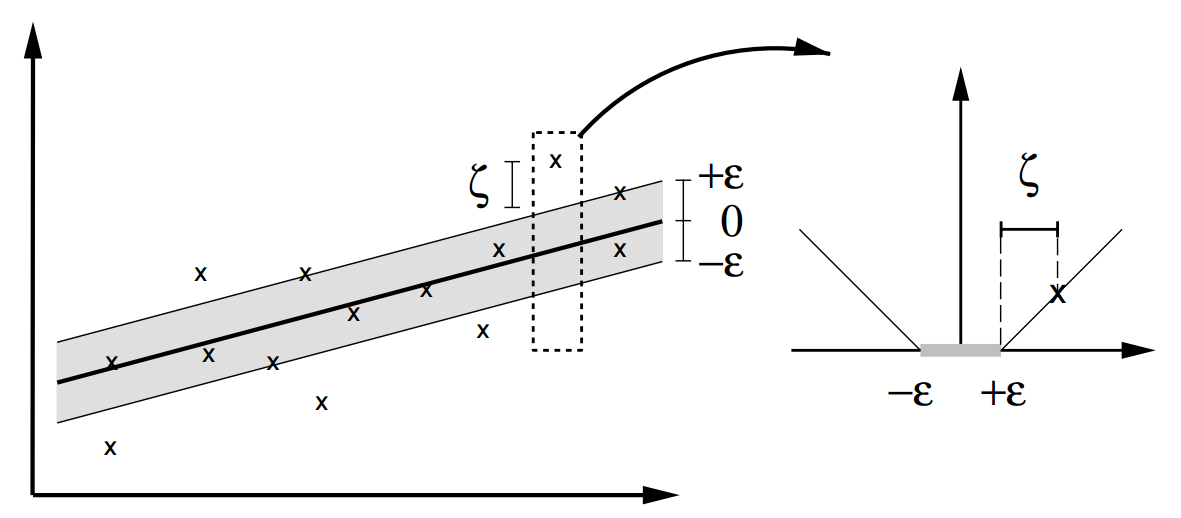}
	\caption[The $\epsilon$-insensitive loss function]{The $\epsilon$-insensitive loss function of the SVR algorithm. Slack variable $\zeta$ captures the loss above the $\epsilon$-tube. Points within the grey shaded area have no impact on the loss. In contrast, all other observations contribute linearly to the loss. Source: \cite{Smola.2004}.}
	\label{loss_fct}
\end{figure}

To measure empirical loss (and thus model fit) using (\ref{Eq:svr_alt_loss_fct}) \cite{Vapnik.1995} reformulates (\ref{svm_mean_eq}) using slack variables $\zeta$ and $\zeta^{*}$ that capture losses above and below the $\epsilon$-tube around $f(x)$, respectively. Figure~\ref{loss_fct} depicts this approach. Only points outside the gray shaded $\epsilon$-tube contribute linearly to the loss function. \\ 

Integrating the slack variables $\zeta$ and $\zeta^*$ into (\ref{svm_mean_eq}), the task to estimate a SVR model is equivalent to solving:

\begin{equation}
\begin{aligned}
& \text{minimize} & & \frac{1}{2} \|\omega\|^2 + C\sum_{i = 1}^{n}(\zeta_i + \zeta^{*}_i) \\
& \text{subject to} & &  \begin{cases}
y_i - \omega^\top x_i - b \le \epsilon + \zeta_i \\
\omega^\top x_i + b - y_i \le \epsilon + \zeta_i^{*} \\
\zeta_i, \zeta_i^{*} \ge 0,
\end{cases} 
\end{aligned}
\label{svr_opt_fct}
\end{equation}
where $C > 0$ is a regularization parameter to balance between model fit and complexity \citep[e.g.,][]{Hastie.2009}. Larger (smaller) values of $C$ put more (less) weight on maximizing model fit during SVR learning. 
 
To capture nonlinear relationships between covariates and the response variable, SVR maps the input data into a higher dimensional feature space. The linear regression is then constructed in the transformed space, which corresponds to a nonlinear regression in the input space. The transformation is feasible from a computational point of view because SVR calculates the mapping by implicitly using a kernel function 
$k(x_i^\top x)=\phi^\top(x_i) \phi(x)$. To implement this approach, it is common practice to estimate a SVR model through solving the dual of (\ref{svr_opt_fct}), which is given as \citep[e.g.,][]{Smola.2004}:

\begin{equation}
\begin{aligned}
& \text{maximize} 
	-\dfrac{1}{2} \sum_{i,j = 1}^{n}(\rho_i - \rho_i^*)(\rho_j - \rho_j^*)x_i^\top x_j 
	-\epsilon\sum_{i = 1}^{n}(\rho_i + \rho_i^*) + \sum_{i = 1}^{n}y_i(\rho_i - \rho_i^*)
 \\
& \text{subject to}  \begin{cases} \sum_{i = 1}^{n} (\rho_i - \rho_i^*) = 0\\
	\rho_i, \rho_i^* \in [0, C].
	\end{cases}
\end{aligned}
\label{dual_lin_opt}
\end{equation}
\newline
The dual program (\ref{dual_lin_opt}) includes the input data only in the form of scalar products $x_i^\top x_j$. Replacing the scalar product by means of a kernel function is thus straightforward and does not affect the solver. In this work, we employ the Gaussian radial basis function (RBF) kernel (\ref{rbf_kernel}) which is defined as
\begin{equation}
\begin{aligned}
k(x_i^\top x)= \exp \left(-\dfrac{\| x - x_i \|^2}{2 \gamma^2} \right)
\end{aligned}
\label{rbf_kernel}
\end{equation}
where the meta-parameter $\gamma>0$ governs the width of the Gaussian function and needs to be set by the modeler. The RBF kernel is used because it includes other kernels as special cases, possesses numerical advantages compared to alternatives, and often performs well in practical applications. Moreover, the RBF kernel can capture nonlinear relations. Other kernels that are usually presented as potential choices are e.g. the linear, polynomial or sigmoid kernel \citep[e.g.,][]{Smola.2004,Hastie.2009}. \cite{Keerthi.2003} show that the linear kernel is a special case of the RBF kernel. Moreover, the polynomial kernel has more parameters than the RBF kernel that make the tuning process more costly. Another advantage of the RBF over the polynomial kernel is that the polynomial kernel can converge to infinity which can cause numerical instability. In contrast, the domain of the RBF kernel is always between 0 and 1. Moreover, \cite{Lin.2003} show that the sigmoid kernel behaves similar to the RBF kernel for certain parameters.

In order to construct the regression function (\ref{Eq:svr_lin_base_mdl}), the weight vector $\omega$ is represented as a linear combination of observations in the training set:

\begin{equation}
\omega = \sum_{i = 1}^{n} (\rho_i - \rho_i^*)x_i.
\label{Eq:weight_lin_svr}
\end{equation}
\newline

More specifically, for observations $x_i$ where $f(x_i)$ is within the $\epsilon$-tube holds that  $\rho_i = \rho_i^* = 0$. Consequently, $f(x)$ depends only on the observations outside the $\epsilon$-tube. These $x_i$ are called support vectors. Accordingly, (\ref{Eq:weight_lin_svr}) is also called the support vector expansion of $\omega$. Rewriting the regression function in terms of the support vector expansion gives the SVR forecasting model:

\begin{equation}
f(x) = \sum_{i = 1}^{n}(\rho_i - \rho_i^*)x_i^\top x + b.  
\label{sv_exp_lin}
\end{equation}
\newline

\noindent In the nonlinear case, the scalar product in (\ref{sv_exp_lin}) is once again replaced by a kernel function.

\subsection{SVR-GARCH-KDE Hybrid}\label{Sec:SVRGARCHKDE}
In the following section, we introduce a nonlinear GARCH hybrid to forecast VaR based on a combination of SVR and KDE. Subsequently, we elaborate on the estimation of the corresponding forecasting model. 

We assume the distribution of the return series $r_t$ to belong to the location-scale class, such that:

\begin{equation}
r_t = \mu_t + u_t = \mu_t + \sigma_t z_t, \qquad z_t \sim (0,1) \ \text{i.i.d.}
\label{Eq:hybrid_locs}
\end{equation}


\noindent Consider an ARMA structure for the mean model where the only assumption  about the error distribution is a zero mean and a finite variance. In addition, recall equation (\ref{Eq:garch_arma2}), which shows that GARCH processes can also be given an ARMA representation. This leads to the following mean and variance model:
\begin{align}
r_t &= c + \sum_{i = 1}^{s} \alpha_i r_{t-i} + \sum_{j = 1}^{d} \kappa_j u_{t-j} + u_t, \quad u_t \sim (0, \sigma_t^2) \\
u_t^2 &=\omega + \sum_{i = 1}^{\max(p,q)} (\delta_i + \beta_i) u_{t-i}^2 - \sum_{j = 1}^{p} \beta_j \nu_{t-j} + \nu_t.
\end{align}

\noindent Let $e = \max(p,q)$, $\mathbf{r}_{t,s} = (r_{t-1}, r_{t-2}, \dots, r_{t-s})$, $\mathbf{u}_{t,k} = (u_{t-1}, u_{t-2}, \dots, u_{t-k})$ and 
$\boldsymbol{\nu}_{t,p} = (\nu_{t-1}, \nu_{t-2}, \dots, \nu_{t-p})$. Then,  following \cite{Chen.2009}, we introduce the nonlinear and nonparametric functions $h$ and $g$ such that the conditional mean and variance models of $r_t$ are

\begin{align}
r_t &= h(\mathbf{r_{t,s}}, \mathbf{u_{t,d}}) + u_t \qquad u_t \sim (0, \sigma_t^2) \label{Eq:garch_mean_no_dis}\\
u_t^2 &= g(\mathbf{u_{t,e}}, \boldsymbol{\nu_{t,p}}) +  \nu_{t} \qquad \nu_t \sim WN(0,a_t^2)\label{Eq:garch_var_no_dis}
\end{align}


\noindent where $ WN(0,a_t^2)$ denotes white noise with expectation zero and variance $a_t^2$. We propose to estimate $h(\cdot)$ and $g(\cdot)$ using SVR. The estimates for $\mu_t$ and $\sigma_t$ in (\ref{Eq:hybrid_locs}) are then obtained as:

\begin{align}
\widehat{\mu}_t &= h(\mathbf{r_{t,s}}, \mathbf{u_{t,d}}) \label{Eq:svr_mean_estimate} \\
\widehat{\sigma}_t &= \sqrt{g(\mathbf{u_{t,e}}, \boldsymbol{\nu_{t,p}})}.
\end{align}

\noindent By defining the estimated residuals as $\widehat{u}_t = r_t - \widehat{\mu}_t$, estimates of $z_t$ are obtained as
\begin{equation}
\widehat{z}_t = \dfrac{\widehat{u}_t}{\widehat{\sigma}_t}.
\label{Eq:standardizes_residuals}
\end{equation}

%
\noindent Then, for $\widehat{Q}_{\widehat{z}}(\alpha)$ being the estimated quantile function of $z$, the VaR estimate for $r_t$ is:

\begin{equation}
\widehat{VaR}_t^\alpha = - \left\{ h(\mathbf{r_{t,s}}, \mathbf{u_{t,d}}) +\sqrt{g(\mathbf{u_{t,e}}, \boldsymbol{\nu_{t,p}})} \widehat{Q}_{\widehat{z}}(\alpha) \right\}.
\label{Eq:nonlin_var}
\end{equation}

\noindent whereby we estimate $\widehat{Q}_{\widehat{z}}(\alpha)$ using KDE.

We now present a procedure to estimate VaR as in (\ref{Eq:nonlin_var}) and describe it in the context of producing one-day-ahead VaR forecasts. A step-by-step overview is given in Algorithm \ref{Algo:svr-garch-kde_algo}. Let $\{r_t\}_{t = 1}^T$ be the training set consisting of the daily returns from a portfolio where $r_T$ is the most recent observation. In the first step, we model the mean process (\ref{Eq:garch_mean_no_dis}). To do this, we estimate an AR($s$) model using SVR to obtain the estimated returns $\{\widehat{r}_t\}_{t = 1+s}^T$. The set of estimated residuals $\{\widehat{u}_t\}_{t = 1 + s}^T$ is derived as $\widehat{u}_t = r_t - \widehat{r}_t$. Then, a moving average (MA) part can be introduced to the model such that $r_t$ can be modeled as an ARMA($s$,$d$) process by running SVR and including $\widehat{u}_t$. The sets of estimated returns and residuals from the ARMA($s$,$d$) model are denoted as $\{\widehat{r}_t^*\}_{t = 1 + s + d}^T$ and $\{\widehat{u}_t^*\}_{t = 1 + s + d}^T$ , respectively.

 We also estimate the variance process in a two step approach and start by fitting the squared mean model residuals $\{\widehat{u}_t^{*2}\}_{t = 1 + s +d}^T$ in the way of an AR($e$) process with SVR. Based on this, fitted variances $\{\widehat{\sigma}_t^2\}_{t = 1 + s+d +e}^T$ are obtained. Then, an ARMA model for (\ref{Eq:garch_var_no_dis}) is obtained in the same way as for the mean process by using the estimated model residuals $\{\widehat{\nu}_t\}_{t = 1 + s + d + e}^T$ where $\widehat{\nu}_t = \widehat{u}_t^{*2} - \widehat{\sigma}_t^2$. Consequently, the final set of fitted variances is denoted as $\{\widehat{\sigma}_t^{*2}\}_{t = 1 + s +d+e+p}^T$. No assumptions are made about the starting values of the residuals. Hence, the final estimation of (\ref{Eq:garch_var_no_dis}) is done using data for $T-s-d-e-p$ time points. Since SVR is applied without introducing further restrictions, it is not ensured that $\widehat{\sigma}_t^2$ and $\widehat{\sigma}_t^{*2}$ are positive. Therefore, if the SVR estimate is $\widehat{\sigma}_t^{\left(*\right)2} \le 0$, it will be replaced by the last positive estimated variance. In case the first fitted variance is negative, it will be replaced by the first squared residual from the final mean model. 

The set of estimated standardized residuals $\{\widehat{z}_t\}_{t = 1 + s +d+e+p}^T$ can be computed by applying (\ref{Eq:standardizes_residuals}). However, $\widehat{z}_t$ does not necessarily have zero mean and unit variance. Hence, we perform the quantile estimation using scaled standardized residuals $\widehat{z}_t^*$:

\begin{equation}
\widehat{z}_t^* = \dfrac{\widehat{z}_t - \overline{\widehat{z}_t}}{\sqrt{\frac{1}{T-1}\sum_{i = 1}^{T}  (\widehat{z}_t - \overline{\widehat{z}_t})^2  }}
\end{equation}

\noindent where $\overline{\widehat{z}_t}$ denotes the empirical mean of $\widehat{z}_t$. The forecasted mean $\widehat{\mu}_{T+1}$ and standard deviation $\widehat{\sigma}_{T+1}$ are obtained from the mean and variance model. Finally, we use KDE to estimate the $\alpha$-quantile of $\{\widehat{z}^*_t\}_{t = 1 + s + d+e+p}^T$. Then, the one-day-ahead VaR forecast is: 


\begin{equation}
\widehat{VaR}_{T+1}^\alpha = -[ \widehat{\mu}_{T+1} + \widehat{\sigma}_{T+1} \widehat{Q}_{\widehat{z}^*} (\alpha)].
\label{Eq:var_est_svrkde}
\end{equation}

An important aspect to note is that SVR and KDE depend on hyperparameters which cannot be derived analytically but must be found computationally. Therefore, the above described estimation procedure describes the estimation process only for one fixed set of hyperparameters. The hyperparameters for SVR given a RBF kernel is used are $\epsilon$, $\gamma$ and $C$. Additionally, if the SVR kernel is not set beforehand, it can be seen as hyperparameter itself. In the context of KDE the bandwidth $h$ and the KDE specific kernel function $K$ are hyperparameters. Since the goal is to train an efficient model on a purely data driven basis it is advisable to derive the hyperparameters computationally. However, for some hyperparameters exist theoretical results that support fixing them beforehand. As stated in Section \ref{Sec:meths_svr} the RBF kernel is a reasonable choice in SVR. Moreover, for KDE the kernel choice has only a low practical relevance and there exist rule-of-thumb estimators for $h$ when a Gaussian kernel  is used which are computationally inexpensive \citep{Hardle.2004, Silverman.1986}. Hence, the most relevant aspect in hyperparameter tuning of the SVR-GARCH-KDE hybrid are $\epsilon$, $\gamma$ and $C$ for the SVR with a RBF kernel. It is important to note that the overall goal is to forecast VaR. Hence, the hyperparameters should be set with respect to the measure that is used to evaluate quantile forecasts. There exist different approaches as grid search, random search or more advanced optimization strategies that can be used to automate the process of hyperparameter optimization.

%

\begin{algorithm}
	\caption{SVR-GRACH-KDE Estimation Algorithm for Forecasting VaR}
	\label{Algo:svr-garch-kde_algo}
	\begin{algorithmic}[1]
		
		\State AR($s$) model for $\{r_t\}_{t=1}^T$ using SVR
		\vspace{0.1cm}
		\State Get errors from Step 1 $\{\widehat{u}\}_{t=1+s}^T$
		\vspace{0.1cm}
		\State ARMA($s$,$d$) model for $\{r_t\}_{t=1+s}^T$ with results from Step 2 using SVR
		\vspace{0.1cm}
		\State Get errors from Step 3 $\{\widehat{u}^*\}_{t=1+s+d}^T$
		\vspace{0.1cm}
		\State AR($e$) model for $\{\widehat{u}_t^{*2}\}_{t=1+s+d}^T$ using SVR
		\vspace{0.1cm}
		\State Get errors from Step 5 $\{\widehat{\nu}_t\}_{t=1+s+d+e}^T$
		\vspace{0.1cm}
		\State ARMA($e$,$p$) model for $\{\widehat{u}_t^{*2}\}_{t=1+s+d+e}^T$  with results from Step 6 using SVR
		\vspace{0.1cm}
		\State Obtain volatility estimates $\{\widehat{\sigma}^*\}_{t=1+s+d+e+p}^T$ from Step 7
		\vspace{0.1cm}
		\State Get standardized residuals $\widehat{z}_t = \widehat{u}_t^* / \widehat{\sigma}_t$ for $t = 1+s+d+e+p, \dots, T$
		\vspace{0.1cm}
		\State Scale $\{\widehat{z_t}\}_{t=1+s+d+e+p}^T$ to zero mean and unit variance and obtain $\{\widehat{z}_t^*\}_{t=1+s+d+e+p}^T$
		\vspace{0.1cm}
		\State Estimate the $\alpha$-quantile $\widehat{Q}_{\widehat{z}^*}(\alpha)$ with KDE
		\vspace{0.1cm}
		\State Obtain $\widehat{r}_{T+1}$ and $\widehat{\sigma}_{T+1}$ by using the models from Step 3 and 7
		\vspace{0.1cm}
		\State VaR forecast: $\widehat{VaR}_{T+1}^\alpha = -[ \widehat{\mu}_{T+1} + \widehat{\sigma}_{T+1} \widehat{Q}_{\widehat{z}^*} (\alpha)]$

	\end{algorithmic}
\end{algorithm}

After defining the SVR-GARCH-KDE hybrid the question arises how model complexity, predictive power and the computational time for making VaR forecasts are related. Regarding the time to compute predictions the KDE part is fixed beforehand. This is can be seen in Equation \ref{Eq:kde_definition} and Algorithm \ref{Algo:svr-garch-kde_algo}. In order to compute an estimate the sum needs to be evaluated making the computational complexity $O(n)$ where $n$ is number of observations. Regarding the SVR-GARCH-KDE hybrid the computational complexity of the KDE part is reduced depending on the order of the autoregressive and moving-average part of the mean and variance process to $O(n-s-d-e-p)$. With respect to SVR the complexity of making a prediction is $O(n_{SV} d)$, where $d$ is the dimension of the feature space and $n_{SV}$ the number of SVs. For the proposed application of forecasting the mean and variance of a financial time series the number of features is usually low. Therefore, the computational time for generating predictions is mainly driven by the sample size and negligible in this application.
	
Regarding the predictive performance, the regression function of the mean and variance process is determined by the SVs which implies that the higher the number of SVs, the higher the complexity. Using a SVR decision function with a high number of SVs can, therefore, lead to a mediocre out-of-sample performance. However, it is not possible to make an ex ante exact statement since the Vapnik-Chervonenkis dimension of the RBF kernel is infinite \citep{Burges.1998}. 
Moreover, the goal of the SVR-GARCH-KDE hybrid is to forecast quantiles. Evaluating the performance ex ante does, therefore, depend on the measure that is used to evaluate forecasted quantiles rather than the statistical properties of KDE and potential error bounds of SVR. In Section \ref{Sec:emp_study} the framework of \cite{Christoffersen.1998} will be applied. For achieving a good performance it is necessary to tune the parameters of the SVR and KDE part appropriately with respect to the target measure. Consequently, in this study the focus is put on the out-of-sample performance to optimize the quality of predictions and the hyperparameter tuning is done with a separate training set.

\section{Empirical Study}\label{Sec:emp_study}

\subsection{General Setting}

The SVR-GARCH-KDE hybrid is tested using stock indices to evaluate the performance for different regions. We consider three indices, namely the Euro STOXX 50, S\&P 500 and Nikkei 225 which represent the Euro zone, the USA and Japan, respectively. The analysis is based on the log-returns of the adjusted index closing prices $P_t$:

\begin{equation}
r_t = \log(P_t) - \log(P_{t-1}).
\end{equation}

\noindent The descriptive statistics of the analyzed indices are given in Table \ref{Tab:desc_stats}.

\begin{table}[ht]
\centering
\begin{tabular}{l|c|c|c|c|c|c|c}
  \hline
  \hline
\textbf{Index} & \textbf{1st Quartile} & \textbf{Mean} & \textbf{Median} & \textbf{3rd Quartile} & \textbf{Variance} & \textbf{Skewness} & \textbf{Kurtosis} \\ 
  \hline
EuroStoxx50 & -0.74 & -0.01 & 0.01 & 0.78 & 2.39 & -0.06 & 5.15 \\ 
  S\&P500 & -0.46 & 0.02 & 0.07 & 0.59 & 1.74 & -0.33 & 9.94 \\ 
  Nikkei225 & -0.76 & 0.00 & 0.05 & 0.88 & 2.68 & -0.51 & 7.49 \\ 
   \hline
    \hline
\end{tabular}
\caption{Descriptive statistics for the log-returns of the analyzed
                      indices in the period from July 1, 2006 to June 30, 2016. Note that the log-returns were multiplied by 100
                      before computing the descriptive statistics.} 
\label{Tab:desc_stats}
\end{table}

We forecast VaR for the quantiles $\alpha \in \{0.01, 0.025, 0.05\}$, considering forecast horizons of one and ten trading days. Estimating VaR for a horizon of ten trading days is especially important regarding the applicability of a VaR model. Besides forecasting VaR for a confidence level of 99\%, which is equivalent to $\alpha = 0.01$ in our setting, a ten days forecast horizon is required in the regulations of the Basel Committee on Banking Supervision. 

In empirical applications, the quality of SVR depends on the kernel and parameter values which need to be set manually. The prevailing approach to determine parameter settings is grid search \citep[e.g.,][]{Lessmann.2017}, which we also apply in this study. For the density estimation via KDE, the Gaussian kernel function in combination with Silverman's rule of thumb are used to reduce computational cost. The Gaussian kernel function in KDE is equivalent to the standard normal distribution:

\begin{equation}
K(u) = \dfrac{1}{2\pi} e^{-\frac{u^2}{2}}.
\label{Eq:kde_gauss}
\end{equation}

\cite{Silverman.1986} showed that when (\ref{Eq:kde_gauss}) is used, robust density estimates can be obtained with the following estimator for the bandwidth $h$:

\begin{equation}
h_{rot} = 0.9 \min\{ \widehat{\sigma}_e, \widehat{\sigma}_{iqr} \} n^{-1/5}.
\label{Eq:kde_silverman}
\end{equation}

In \ref{Eq:kde_silverman} $\widehat{\sigma}_e$ denotes the empirical standard deviation and $\widehat{\sigma}_{iqr}$ is an estimate of the standard deviation that is based on the interquartile range $R$:

\begin{equation}
\widehat{\sigma}_{iqr} = \dfrac{R}{1.34}.
\label{Eq:kde_iqr_sd}
\end{equation}

The rule-of-thumb estimator $h_{rot}$ is computationally inexpensive whereas other proposed approaches to bandwidth estimation are more expensive since they are usually based on cross validation \citep{Hardle.2004}. Moreover, the importance of the kernel choice regarding the performance of KDE is limited. For instance, \cite{Hardle.2004} conclude that the kernel choice has almost no practical relevance after deriving the asymptotic mean integrated squared error for different kernel choices.

The evaluation of the models is based on \cite{Christoffersen.1998} who proposes a likelihood ratio (LR) test framework, which assesses the unconditional and conditional coverage as well as the independence of VaR exceedances. Moreover, \cite{Christoffersen.1998} shows that the test statistic for conditional coverage can be derived as the sum of the test statistics of the test for unconditional coverage and independence of VaR exceedances. Hence, it is possible to test whether the performance of a VaR model in terms of conditional coverage is determined by its ability to achieve correct unconditional coverage or adjust for changing volatility. This is useful in situations where the model has relatively bad conditional coverage but only one of the test statistics for unconditional coverage or independence is small. The hypotheses for the three tests are shown in Table \ref{Tab:hypotheses_cov} where $\alpha$ is the target quantile and $\mathcal{F}_{t-1}$ the information set available at $t-1$. $V_t$ is a series of VaR violations with $V_t = \mathbb{I}(r_t < -VaR_t^{\alpha})$ where $\mathbb{I}(x < c)$ denotes the indicator function:
	\begin{equation}
	\mathbb{I}(x < c) = \begin{cases}
	1 \text{ \ if \ } x < c \\
	0 \text{ \ if \ } x \ge c. \\
	\end{cases} 
	\end{equation}
For the test of independence of violations $V_t$ is assumed to be a binary first-order Markov chain. The corresponding transition probability matrix of $V_t$ is
	
	\begin{equation}
	\Pi_1 = \begin{bmatrix}
	1 - \pi_{01} & \pi_{01} \\
	1 - \pi_{11} & \pi_{11} 
	\end{bmatrix}
	\end{equation}
	
\noindent where 
	
	\begin{equation}
	\pi_{ij} = \mathbb{P}(V_t = j | V_{t-1} = i).
	\end{equation}

\begin{table}[ht]
	\centering
	\begin{tabular}{l|l|l}
		\hline
		\textbf{Test} & \boldmath{$\textbf{H}_0$} & \boldmath{$\textbf{H}_1$} \\ 
		\hline
		\hline
		Unconditional Coverage & $\mathbb{E}[V_t] = \alpha$ & $\mathbb{E}[V_t] \ne \alpha$  \\ 
		\hline
		Independence of Violations & $\pi_{01} = \pi_{11}$ & $\pi_{01} \ne \pi_{11}$ \\ 
		\hline
		Conditional Coverage & $\mathbb{E}[V_t| \mathcal{F}_{t-1}] = \alpha$ & $\mathbb{E}[V_t| \mathcal{F}_{t-1}] \ne \alpha$  \\ 

		\hline
		\hline
	\end{tabular}
	\caption{Hypotheses for evaluating the appropriateness of VaR forecasts with the testing framework introduced by \cite{Christoffersen.1998}.
	\label{Tab:hypotheses_cov}} 
\end{table}

 As criterion for selecting a model from grid search and evaluate the performance of the SVR-GARCH-KDE hybrid with respect to benchmark models, the $p$-value of the test for conditional coverage is used. Since the null hypothesis corresponds to correct conditional coverage which is the desired property, the one with the highest $p$-value is considered to be the best. 

By using the framework of \cite{Christoffersen.1998} to evaluate and select models, the main focus is put on the statistical properties and VaR violations of the considered models. In order to measure the performance from the perspective of a loss function that also takes into account the magnitude of violations, the quadratic and asymmetric loss function that was introduced in one of the seminal papers for evaluating VaR models by \cite{Lopez.1998} will be used: 

\begin{equation}
L(r_t, VaR_t^{\alpha}) = \begin{cases}
1 + (r_t - VaR_t^{\alpha})^2  &\text{ \ if \ } r_t < VaR_t^{\alpha}  \\
0 &\text{ \ if \ } r_t \ge VaR_t^{\alpha}. \\
\end{cases} 
\end{equation}
\color{red}
\noindent The model losses will be evaluated by employing the superior predictive ability test (SPA) framework proposed by \cite{Hansen.2005}. Following the notation of \cite{Hansen.2005}, for a finite set of decision rules $[\delta_{k, t-h}, k = 0, 1, \dots, m]$ made $h$ periods in advance with respect to a random variable $\xi_t$, the relative performance corresponding to the benchmark $\delta_{0, t-h}$ is measured as

\begin{equation}
d_{k,t} = L(\xi_t, \delta_{0, t-h}) - L(\xi_t, \delta_{k, t-h}).
\end{equation}

\noindent Based on the assumption that the alternatives are superior if and only if  $\mathbb{E}[d_{k,t}] > 0$, \cite{Hansen.2005} formulates the hypothesis of interest for $ \boldsymbol{d_t} = (d_{1,t}, \dots, d_{m,t})^\top $ as

\begin{equation}
	 \text{H}_0: \mathbb{E}[\boldsymbol{d_t}] \le \boldmath{0}.
\end{equation}

\noindent A high $p$-value indicates that none of the alternatives is superior to the benchmark. For further information regarding the SPA test we refer to \cite{Hansen.2005}. As for the LR test of \cite{Christoffersen.1998}, the SPA test will be performed for every model. This is done by using every considered model once as benchmark in terms of \cite{Hansen.2005} and comparing it to all other models. Note that due to performing multiple statistical tests the $p$-values should be interpreted rather as an indication of model performance than in the context of a fixed significance level. 

\color{black}

We perform all analyses using the statistical software R. The data has been downloaded from Yahoo Finance using the quantmod package. For SVR, we use the package e1071, which is the R implementation of the LIBSVM library of \cite{Chang.2011}. To reduce computational time, we employ the doParallel package for parallelization of computations. The benchmark methods introduced below are implemented by using the rugarch package. All codes are available on \url{www.quantlet.de}. For details we refer to \cite{Borke.2018} and \cite{Borke.2017}.

\subsection{Benchmark Methods}

To test the SVR-GARCH-KDE hybrid, we compare its performance to the standard GARCH model and two of its variations. In particular, \cite{Franke.2011} state that the most important variations are the EGARCH and TGARCH model. Hence, they serve as benchmarks in the empirical comparison. The EGARCH and TGARCH models are introduced by \cite{Nelson.1991} and \cite{Zakoian.1994}, respectively. In contrast to standard GARCH models, both can account for asymmetric behavior with respect to past positive or negative returns. The two main differences between EGARCH and TGARCH models are that the former has a multiplicative and the latter an additive model structure. Moreover, TGARCH models allow for different coefficients depending on the lags whereas EGARCH models capture the asymmetric behavior for all lags with one coefficient. The GARCH-type models that \cite{Kuester.2005} analyze are coupled with different error distributions, i.e. the normal distribution, $t$-distribution and skewed $t$-distribution. We adopt this approach, which implies that we compare the SVR-GARCH-KDE hybrid to nine benchmarks.

\subsection{Results}

\subsubsection{Model Setting and Tuning}

We assume the mean process of $r_t$ in (\ref{Eq:svr_mean_estimate}) is zero. Moreover, we assume the variance process to have one AR and one MA part. These assumptions are imposed on both the SVR-GARCH-KDE hybrid and the benchmark models. We then forecast VaR  for every index trading day from 2011-07-01 until 2016-06-30. The data is scaled to zero mean and unit variance in the SVR step; as suggested in the documentation of the e1071 package. 

The tuning of the hyperparameters is done in a moving window approach using 251 return observations, which corresponds to approximately one trading year. The hyperparameters are tuned for every combination of index and quantile separately. For a given set of hyperparameters the model is trained based on the returns of the last trading year to predict the next day's VaR. Then, the window is shifted by one trading day. The tuning period reaches from 2006-07-01 until 2011-06-30, where 2006-07-01 marks the date of the first VaR prediction. The considered parameter values in the grid for SVR are

\begin{itemize}
	\item $C \in \{10^{-4}, 10^{-3}, \dots, 10^4 \}$
	\item $\psi \in \{0, 0.1, \dots, 0.9 \}$  where $\epsilon = Q_{u^2_{scale}}(\psi)$
	\item $\gamma \in \{10^{-4}, 10^{-3}, \dots, 10^4 \}$.
\end{itemize}    
 
Note that the second point indicates that tuning is not done over fixed values of $\epsilon$. Instead, in every step of the estimation, $\epsilon$ is determined based on the $\psi$-quantile of the squared scaled disturbances of the mean model. This corresponds to the squared scaled returns because we assume zero mean of $r_t$. The motivation behind this is the tendency of returns to form volatility clusters. Hence, a fixed $\epsilon$ can lead to good results in one volatility regime but might have a poor performance after a regime change. For instance, in the case of a financial crisis, the right tail of the distribution of past volatilities gets thicker. Hence, an $\epsilon$ that depends on the quantile of the distribution will increase such that large volatilities have automatically a higher influence on the estimated parameters from the SVR optimization. By using squared values it is ensured that only positive values are obtained for $\epsilon$. However, notice that the distribution of the scaled squared disturbances, which are used in the SVR training, is shifted to the left of the distribution of the squared scaled disturbances. Hence, it is possible that for high values of $\psi$ no observations are outside the $\epsilon$-tube such that the model cannot be estimated.    

The parameter settings that resulted in the best models for the SVR-GARCH-KDE hybrid during the tuning period are shown in Table \ref{Tab:tuning_case1}.  
It can be seen that especially for $\psi$ only a certain range appears among the best models. Moreover, the optimal $\psi$ tends to be higher for lower quantiles. Based on the obtained parameters, VaR forecasts are produced from 2011-07-01 until 2016-06-30. 


\begin{table}[h!t]
\centering
\begin{tabular}{c|c|c|c|c|c|r|r|r}
  \hline
  \hline
\textbf{C} & $\boldsymbol{\psi}$ & $\boldsymbol{\gamma}$ & \textbf{Index} & \textbf{Quantile} & \textbf{Violations} & \textbf{UC} & \textbf{ID} & \textbf{CC} \\ 
  \hline
  10 & 0.7 & 0.1 & S\&P500 & 1.0 & 0.95 & 86.62 & 63.08 & 87.84 \\ 
  10 & 0.7 & 0.01 & S\&P500 & 2.5 & 2.46 & 93.15 & 96.55 & 96.20 \\ 
  10 & 0.6 & 0.001 & S\&P500 & 5.0 & 5.00 & 99.48 & 99.57 & 99.57 \\ 
  \hline
  100 & 0.8 & 0.01 & Nikkei225 & 1.0 & 0.82 & 50.63 & 68.48 & 73.85 \\ 
  100 & 0.7 & 0.01 & Nikkei225 & 2.5 & 2.70 & 66.43 & 57.80 & 52.61 \\ 
  100 & 0.7 & 0.10 & Nikkei225 & 5.0 & 4.49 & 40.84 & 94.38 & 67.06 \\ 
      \hline
	100 & 0.7 & 0.01 & EuroStoxx50 & 1.0 & 1.02 & 93.28 & 60.41 & 87.11 \\ 
  0.1 & 0.6 & 0.001 & EuroStoxx50 & 2.5 & 2.52 & 96.42 & 97.76 & 97.66 \\ 
  10000 & 0.6 & 10 & EuroStoxx50 & 5.0 & 4.96 & 94.86 & 99.71 & 99.50 \\ 
   \hline
\hline
\end{tabular}
\caption[Grid search results for SVR parameter tuning]{The best models in the tuning period according to the p-value of the test for conditional coverage.
                      UC, ID and CC indicate the p-value of the corresponding LR test. All values in the columns Quantile, Violations,
                      UC, ID and CC are given in percent.} 
\label{Tab:tuning_case1}
\end{table}

\subsubsection{Model Comparison}

\noindent The results are presented for each quantile separately at the end of the section. In the Tables \ref{Tab:case1_1}, \ref{Tab:case1_2.5} and \ref{Tab:case1_5} are the results for the one-day-ahead forecast. The results for the ten-days-ahead forecasts can be found in the Tables \ref{Tab:case1_10days_1}, \ref{Tab:case1_10days_2.5} and \ref{Tab:case1_10days_5}. To clearly identify the best performing models, every table is sorted in descending order for every index according to the $p$-value of the conditional coverage test. The $p$-value of the conditional coverage test is used for sorting to focus rather on statistical properties than pure losses. The abbreviations NORM, STD and SSTD indicate the normal, $t$- and skewed $t$-distribution, respectively. Additionally, the column headers UC, ID, CC and SPA refer to the corresponding $p$-value of the test for unconditional coverage, independence of violations, conditional coverage and superior predictive ability.  

\begin{figure}[h]
	\centering
	\includegraphics[scale = 0.7]{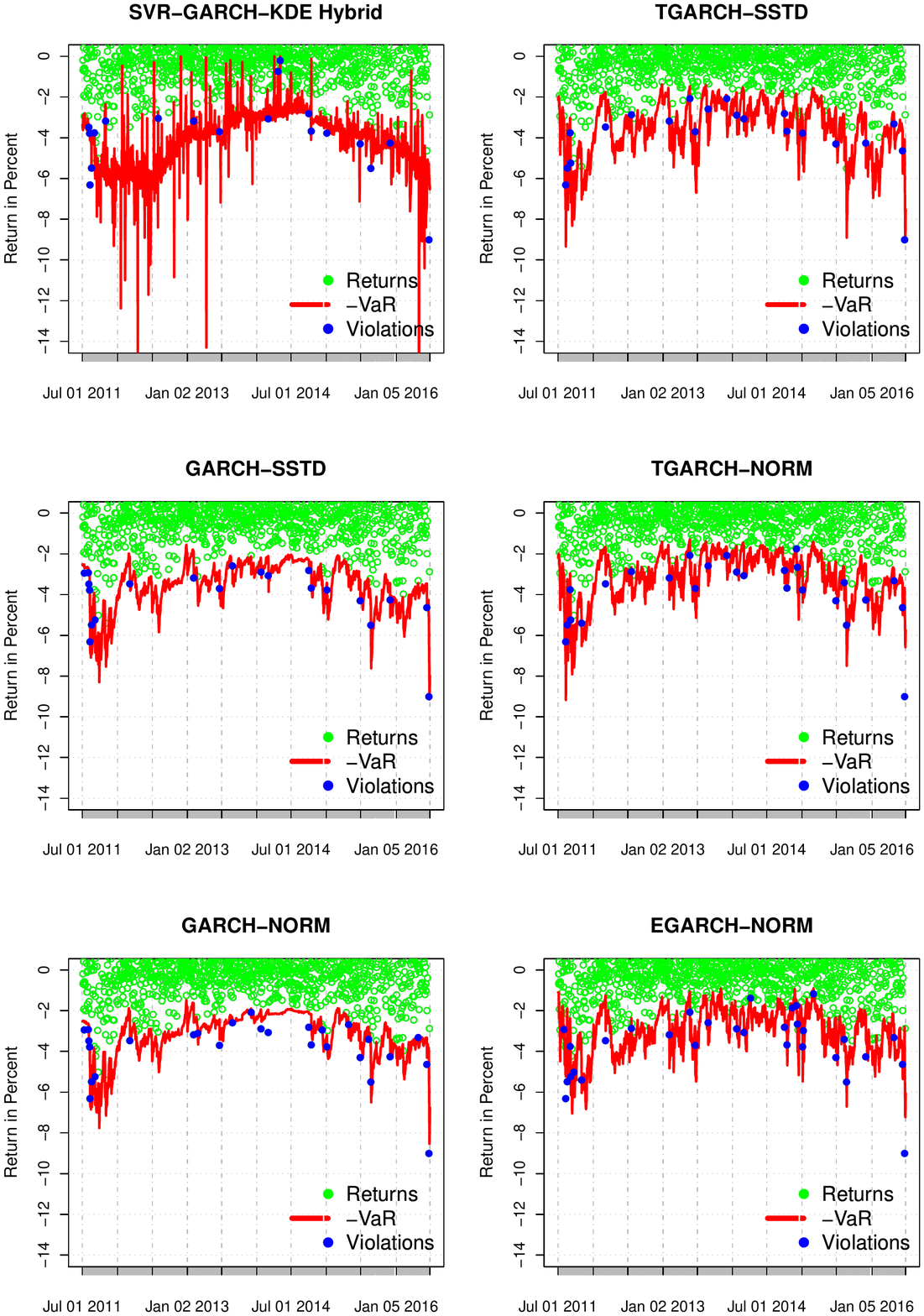}
	\caption{VaR one-day-ahead forecast model comparison for the Euro Stoxx 50 at $\alpha = 0.01$ in the period from July 1, 2011 to June 30, 2016. The SVR-GARCH-KDE hybrid is compared to models that are overall on average better and to the models using a normal distribution.}
	\label{Fig:model_comparision_estoxx_1}	
\end{figure}

\paragraph{One-Step-Ahead Forecast Model Evaluation for $\alpha = 0.01$} The SVR-GARCH-KDE hybrid is the best model for the Euro STOXX 50 for $\alpha = 0.01$. A visualization of its performance is given in Figure~\ref{Fig:model_comparision_estoxx_1}. Here, the SVR-GARCH-KDE hybrid estimates in general higher values for VaR than the other models and exhibits more variability. For the S\&P 500 and Nikkei 225 the SVR-GARCH-KDE hybrid outperforms all models that are coupled with a normal distribution. However, all models using a skewed $t$-distribution perform better. Especially for the Nikkei 225 this is caused by having low unconditional coverage due to risk overestimation. However, since risk overestimation causes less losses the SPA test $p$-value is highest. In general, the models coupled with a normal distribution perform poorly for all indices. This comes as no surprise since the distribution of asset returns is usually leptokurtic.

\begin{figure}[h]
	\centering
	\includegraphics[scale = 0.7]{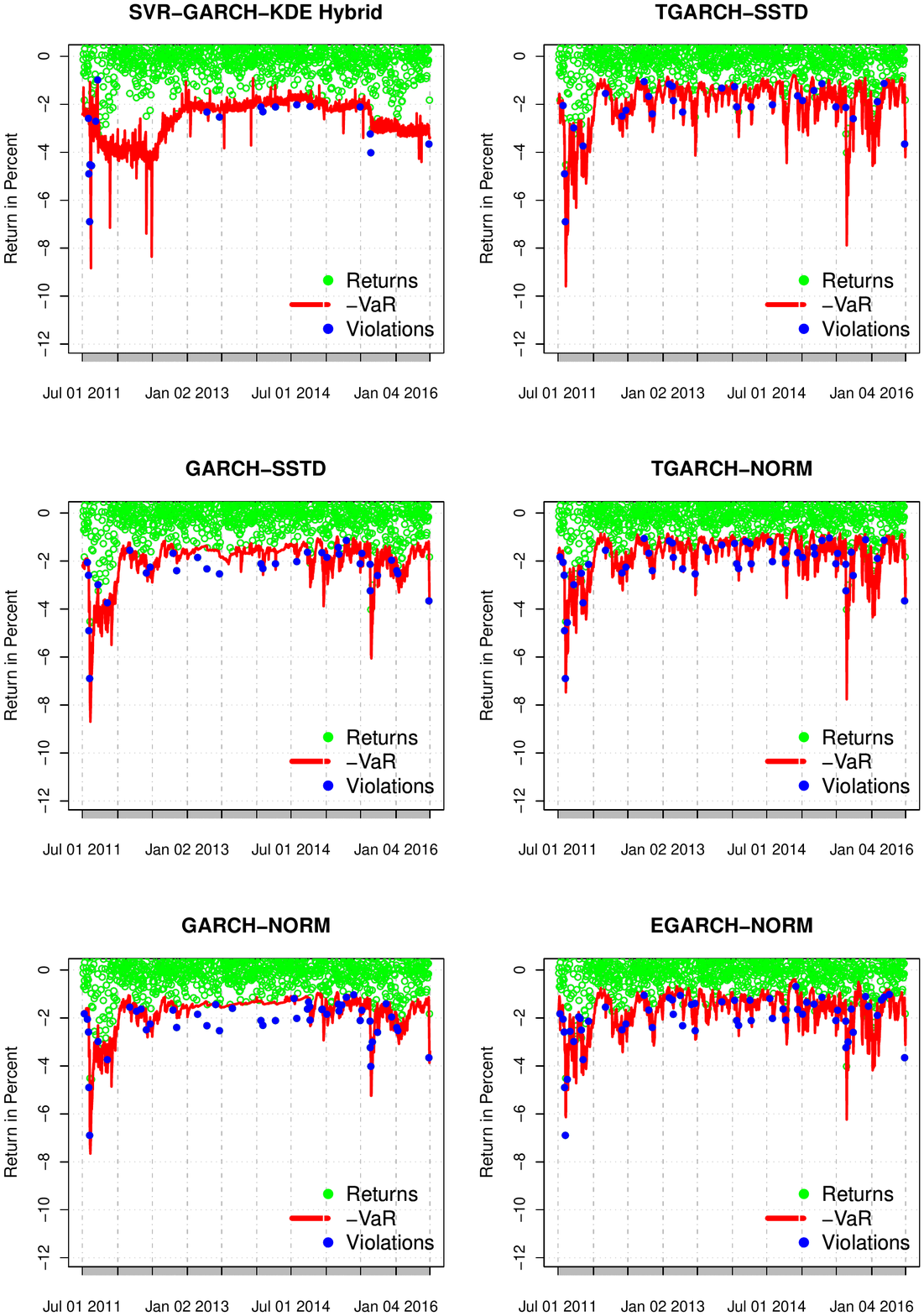}
	\caption{VaR one-day-ahead forecast model comparison for the S\&P 500 at $\alpha = 0.025$ in the period from July 1, 2011 to June 30, 2016. The SVR-GARCH-KDE hybrid is compared to models that are overall on average better and to the models using a normal distribution.}
	\label{Fig:model_comparision_SP500_25}	
\end{figure}

\paragraph{One-Step-Ahead Forecast Model Evaluation for $\alpha = 0.025$} The performance of the SVR-GARCH-KDE hybrid at the 2.5\% level is not as good as for $\alpha = 0.01$ for the Euro STOXX 50. We observe the lowest $p$-value in the test for independence of violations but the third best regarding the test for unconditional coverage. Interestingly, although the TGARCH and EGARCH model account for asymmetries in volatility, the former is the best and the latter the worst variance model. A relatively high risk overestimation for the S\&P 500 and Nikkei 225 causes the SVR-GARCH-KDE hybrid to be on the sixth and ninth rank, respectively. As seen for $\alpha = 0.01$ this leads, however, to high $p$-values in the SPA test due to lower losses. The performance for the S\&P 500 can be seen in Figure~\ref{Fig:model_comparision_SP500_25}.

\begin{figure}[h]
	\centering
	\includegraphics[scale = 0.7]{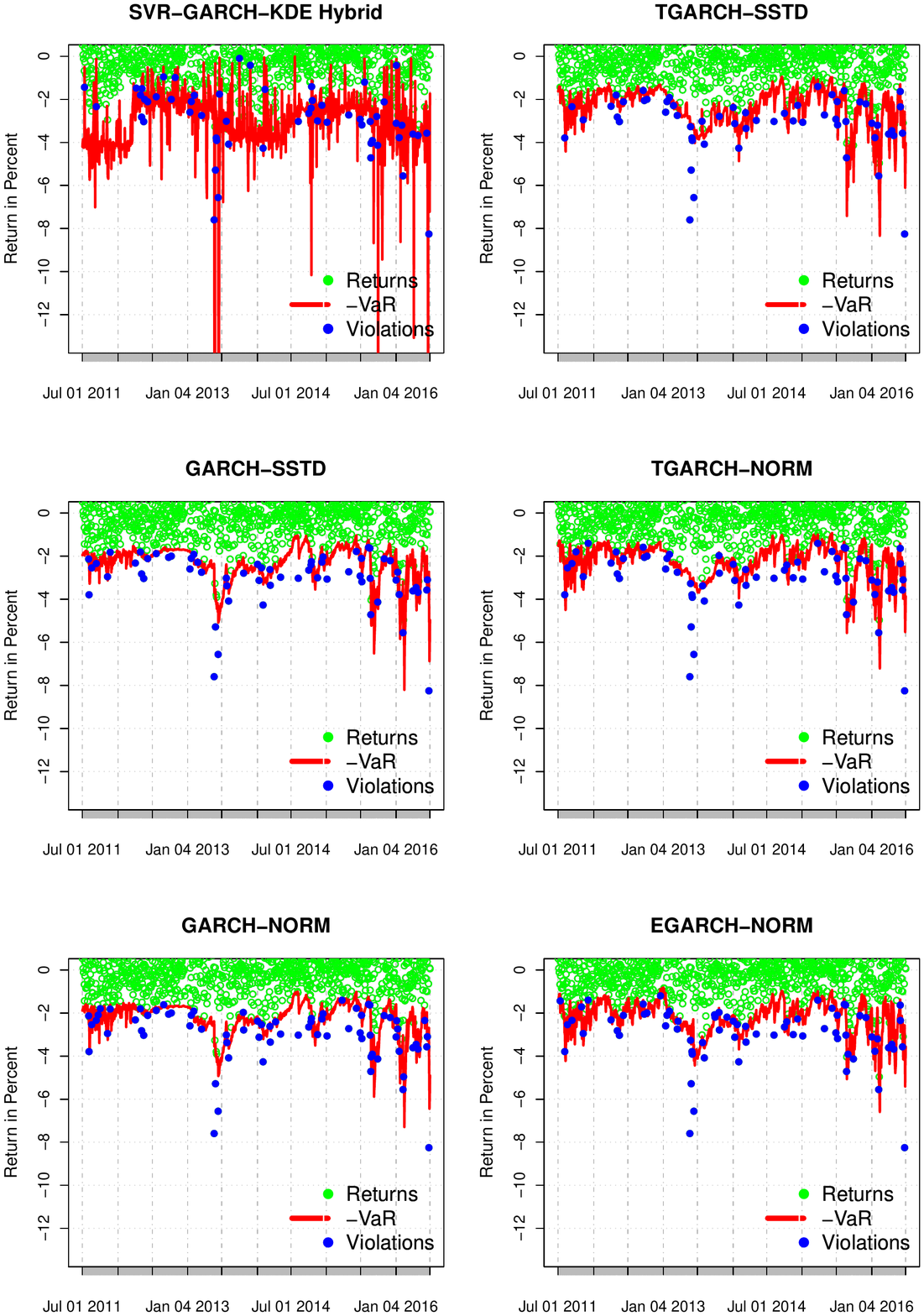}
	\caption{VaR one-day-ahead forecast model comparison for the Nikkei 225 at $\alpha = 0.05$ in the period from July 1, 2011 to June 30, 2016. The SVR-GARCH-KDE hybrid is compared to models that are overall on average better and to the models using a normal distribution.}
	\label{Fig:model_comparision_nikkei_5}	
\end{figure}

\paragraph{One-Step-Ahead Forecast Model Evaluation for $\alpha = 0.05$} The best performance of the SVR-GARCH-KDE hybrid for $\alpha= 0.05$ is rank two for the Nikkei 225. Here, it is only beaten by the EGARCH-SSTD model. A visualization is given in Figure~\ref{Fig:model_comparision_nikkei_5}. For the other indices the SVR-GARCH-KDE hybrid ranks on place five. This is caused by having relatively low $p$-values for the ID test. In terms of UC, the SVR-GARCH-KDE hybrid is the second and third best model for the S\&P 500 and Nikkei 225, respectively. In comparison to the results for $\alpha = 0.01$, the models with a normal distribution show a better performance for $\alpha \in \{0.025, 0.05\}$. However, using the skewed $t$-distribution leads also for $\alpha \in \{0.025, 0.05\}$ to the best rankings.  

\begin{figure}[h]
	\centering
	\includegraphics[scale = 0.7]{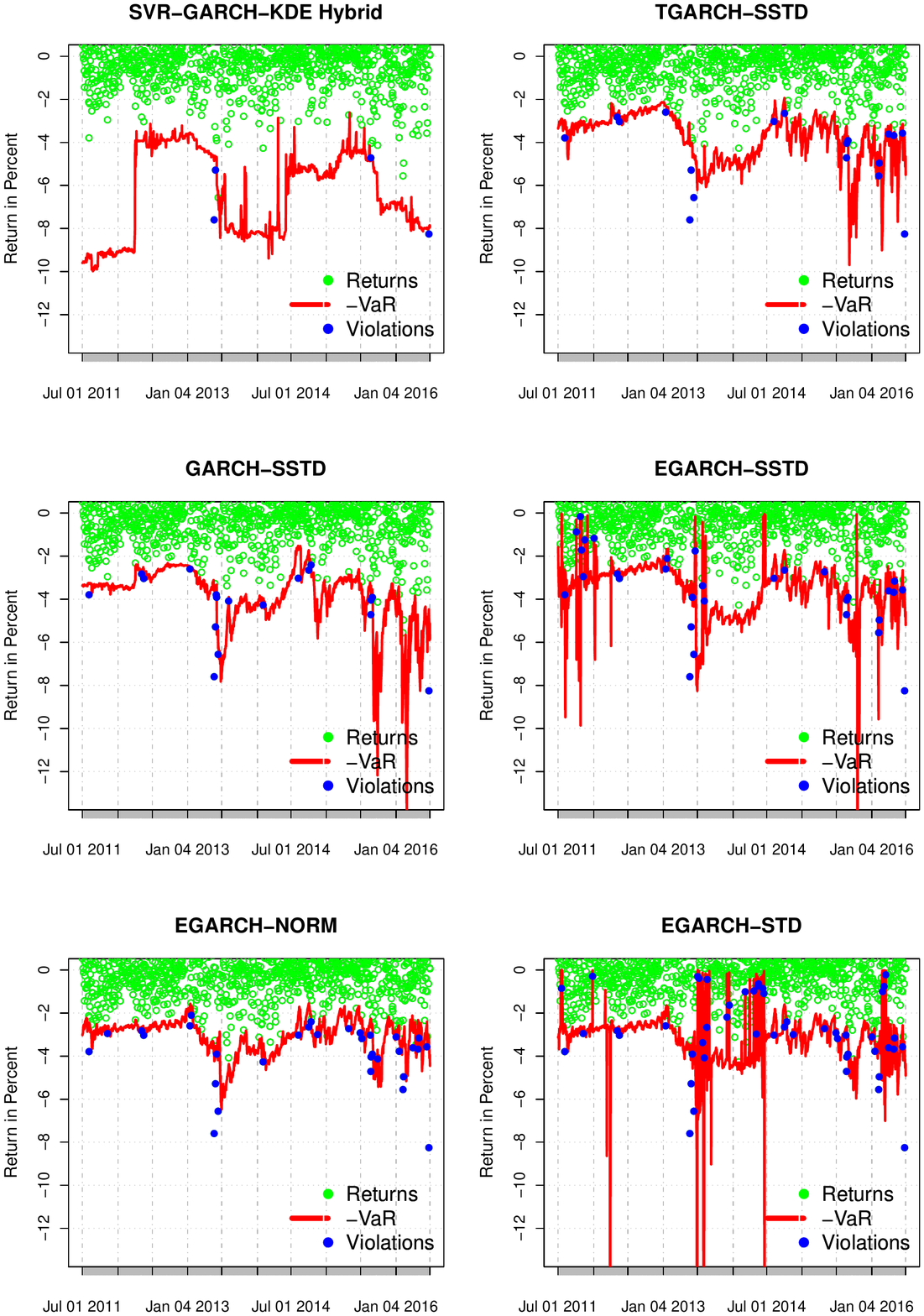}
	\caption{VaR ten-days-ahead forecast model comparison for the Nikkei 225 at $\alpha = 0.01$ in the period from July 1, 2011 to June 30, 2016. The SVR-GARCH-KDE hybrid is compared to models that are overall on average better and to the EGARCH models.}
	\label{Fig:model_comparision_nikkei_1_10days}	
\end{figure}

\paragraph{Ten-Steps-Ahead Forecast Model Evaluation for $\alpha = 0.01$} As for the one-step-ahead forecasts, the SVR-GARCH-KDE hybrid is also the best model for the Euro STOXX 50. For the S\&P 500 and Nikkei 225 the SVR-GARCH-KDE hybrid is the second and third best model, respectively. For all indices, it is the only model not underestimating the risk. Especially the EGARCH models perform poorly. A visualization is given in Figure~\ref{Fig:model_comparision_nikkei_1_10days}. In contrast to the one-step-ahead forecasts, the models using a normal distribution are not shown but the EGARCH models due to their bad performance for ten-days-ahead forecasts. With respect to the SPA test, the SVR-GARCH-KDE hybrid causes less severe losses than any other model for all levels of the ten-steps-ahead forecast horizon.

\begin{figure}[h]
	\centering
	\includegraphics[scale = 0.7]{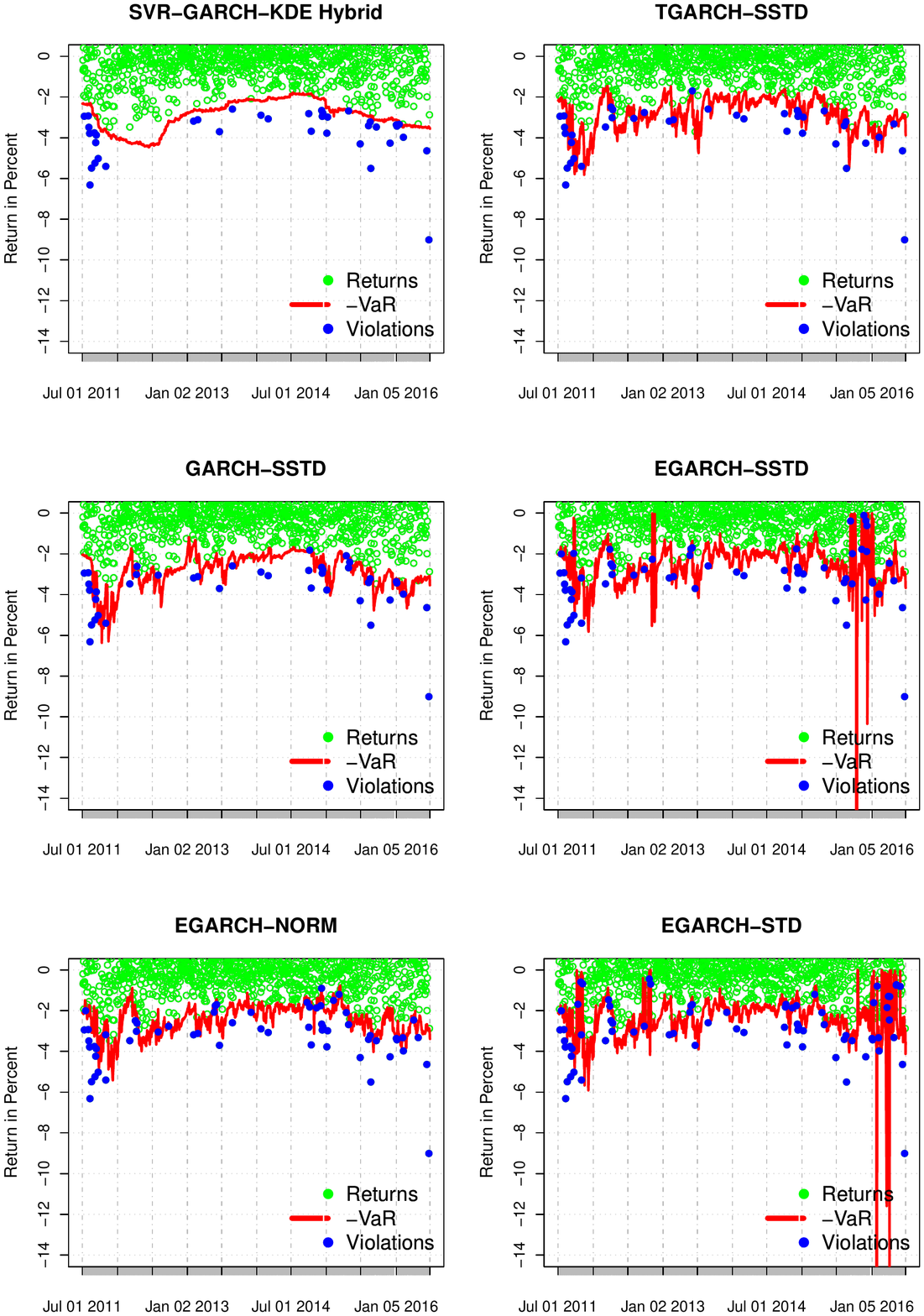}
	\caption{VaR ten-days-ahead forecast model comparison for the Euro Stoxx 50 at $\alpha = 0.025$ in the period from July 1, 2011 to June 30, 2016. The SVR-GARCH-KDE hybrid is compared to models that are overall on average better and to the EGARCH models.}
	\label{Fig:model_comparision_eurostoxx_25_10days}	
\end{figure}

\paragraph{Ten-Steps-Ahead Forecast Model Evaluation for $\alpha = 0.025$} Overall, the SVR-GARCH-KDE hybrid performs worse at the 2.5\% than for the 1\% level. In case of the the S\&P 500 and Nikkei 225 this is mainly driven by overestimating the risk leading to low $p$-values in the UC test. In contrast, even though the SVR-GARCH-KDE hybrid has the highest $p$-value in the UC test for the EURO STOXX 50, it is relatively low for the ID test. An exemplary visualization is given for the EURO STOXX 50 in Figure~\ref{Fig:model_comparision_eurostoxx_25_10days}.

\begin{figure}[h]
	\centering
	\includegraphics[scale = 0.7]{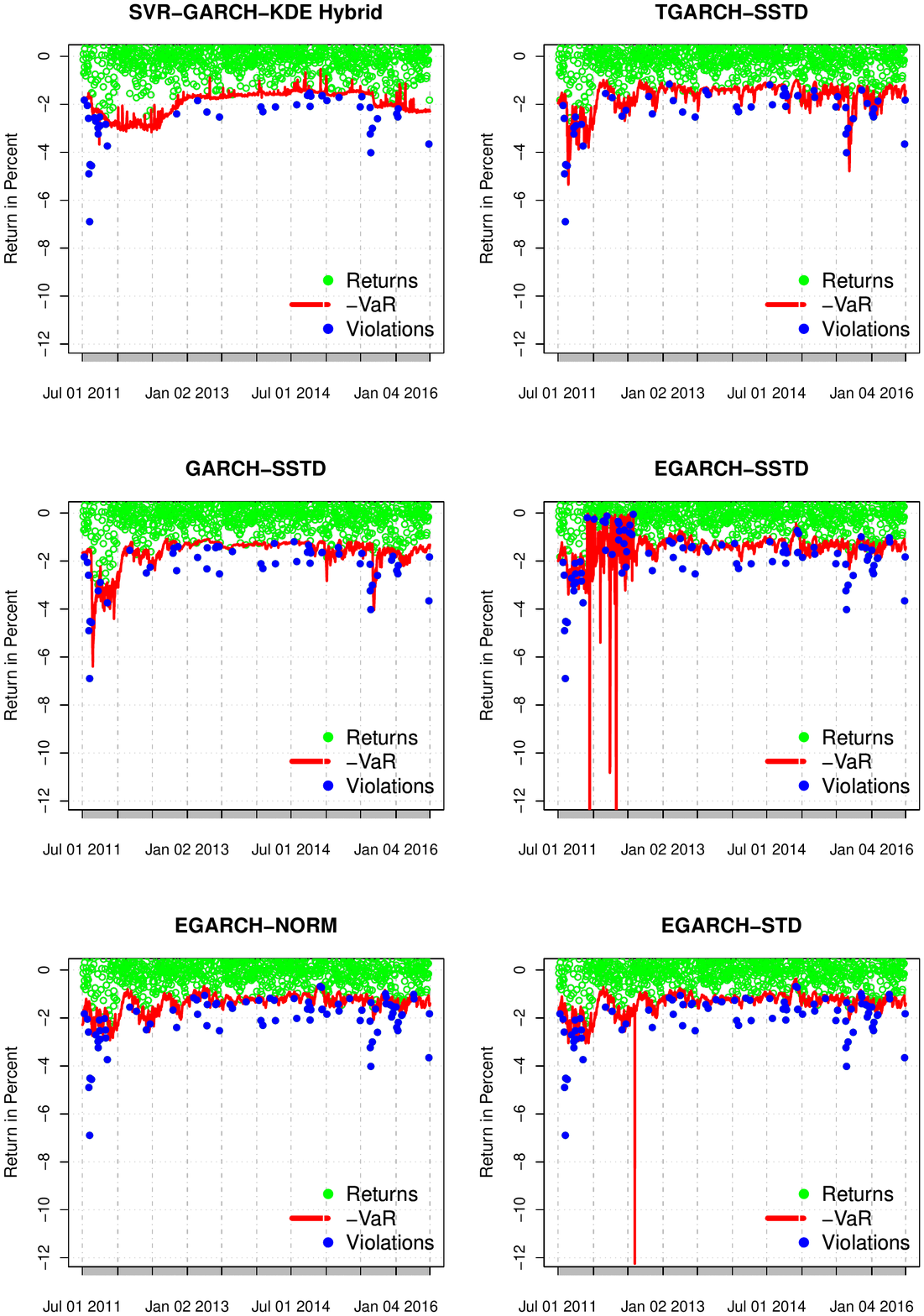}
	\caption{VaR ten-days-ahead forecast model comparison for the S\&P 500 at $\alpha = 0.05$ in the period from July 1, 2011 to June 30, 2016. The SVR-GARCH-KDE hybrid is compared to models that are overall on average better and to the EGARCH models.}
	\label{Fig:model_comparision_sap_5_10days}	
\end{figure}

\paragraph{Ten-Steps-Ahead Forecast Model Evaluation for $\alpha = 0.05$} The performance of the SVR-GARCH-KDE hybrid for $\alpha= 0.05$ is mixed. While it is by far the best model for the EURO STOXX 50, it achieves only rank seven and nine for the Nikkei 225 and S\&P 500, respectively. As for $\alpha = 0.025$ this is mainly driven by risk overestimation and low $p$-values for the UC test. Similar to the one-day-ahead forecasts, the models using a normal distribution perform better for $\alpha = 0.05$ than for the other quantiles. The TGARCH-NORM model is even the best for the S\&P 500. A visualization is shown in Figure~\ref{Fig:model_comparision_sap_5_10days}.

\paragraph{Evaluation Summary} Summarizing the results observed across all indices, quantiles and the two forecast horizons, we conclude that the SVR-GARCH-KDE hybrid displays a competitive performance. This can be seen in Table \ref{Tab:mean_ranks} where the mean ranks per index and quantile are presented for each model with respect to the CC and SPA test as well as the two forecast horizons. The SVR-GARCH-KDE hybrid is overall and for the CC test $p$-values the third best model for both forecast horizons. Regarding the SPA test it has rank two for the one-day-ahead forecast horizon and is the best model for the ten-days-ahead forecast horizon. Benchmark models coupled with a normal distribution are usually outperformed by the SVR-GARCH-KDE hybrid. Consequently, the SVR-GARCH-KDE hybrid has a relatively high accuracy especially compared to models with a normal distribution. The importance of this result stems from the fact that GARCH models with a normal distribution can be seen as a popular standard approach for modeling VaR or volatility and are, therefore, a natural benchmark. Additionally, excepting the case of the ten-days-ahead S\&P 500 forecast for $\alpha = 0.05$ there is no setting where the models using a normal distribution perform best. This provides further evidence that using the normal distribution to measure market risk is inappropriate. However, models with a skewed $t$-distribution show in many cases the best performance. For instance, the TGARCH model coupled with the skewed $t$-distribution is with two exceptions always among the top three in terms of the CC test. This confirms the results of previous research showing usually skewed return distributions. Unlike the benchmarks, the SVR-GARCH-KDE hybrid tends to overestimate market risk. This might come from the choice of time interval for SVR parameter tuning. In particular, the tuning period covers the financial crisis of 2008 where market risk was extremely high. However, in the context of risk management, risk underestimation is more critical than risk overestimation because it can lead to bankruptcy in the short term. For instance, assume a hypothetical situation with the goal to forecast the 5\% VaR, where the SVR-GARCH-KDE hybrid and a benchmark model have the same $p$-value regarding the independence test. Additionally, assume the $p$-value of the benchmark in the test of conditional coverage is higher because it has an unconditional coverage of 5.5\% whereas that of the SVR-GARCH-KDE hybrid is 4\%. The 1\% overestimation of the SVR-GARCH-KDE hybrid works like a buffer for model risk since all estimation techniques exhibit statistical uncertainty. Hence, the use of the SVR-GARCH-KDE hybrid may be still more appealing than the use of benchmark models that tend to underestimate risk.



\begin{table}[ht]
	\centering
	\begin{tabular}{l|c|c|c|c|c|c}
		\hline
		 \textbf{Model} & \textbf{Overall} & \textbf{CC 1 Day}  & \textbf{SPA 1 Day}  & \textbf{CC 10 Days} & \textbf{SPA 10 Days}  \\ 
		 \hline
 		\hline
TGARCH-SSTD & 2.3 & 2.2 & 1.4 & 2.3 & 3.2 \\ 
GARCH-SSTD & 2.6 & 2.3 & 2.6 & 2.7 & 2.7 \\ 
SVR-GARCH-KDE & 3.2 & 4.9 & 2.2 & 4.8 & 1.0 \\ 
GARCH-STD & 5.9 & 6.1 & 6.4 & 4.9 & 6.3 \\ 
TGARCH-NORM & 6.0 & 6.0 & 7.7 & 5.1 & 5.1 \\ 
GARCH-NORM & 6.6 & 6.1 & 6.8 & 6.3 & 7.3 \\ 
TGARCH-STD & 6.7 & 5.3 & 9.0 & 5.1 & 7.4 \\ 
EGARCH-SSTD & 6.9 & 5.0 & 5.3 & 9.4 & 7.9 \\ 
EGARCH-NORM & 8.6 & 8.7 & 7.3 & 9.3 & 9.0 \\ 
EGARCH-STD & 8.6 & 8.8 & 7.2 & 9.6 & 8.8 \\ 
		 \hline
		\hline
	\end{tabular}
	\caption[Model comparison via mean ranks]{The models are presented with their mean rank. The mean rank was computed per index and quantile using the ties method \textit{max} in the \textit{frankv} function of the R package data.table. The lower the mean rank, the better the model.}
	\label{Tab:mean_ranks}
\end{table}

\begin{table}[ht]
\centering
\begingroup\fontsize{11pt}{12pt}\selectfont
\begin{tabular}{l|c|c|r|r|r|r}
  \hline
\textbf{Model} & \textbf{Index} & \textbf{Violations} & \textbf{SPA} & \textbf{UC} & \textbf{ID} & \textbf{CC} \\ 
  \hline
\hline
EGARCH-SSTD & S\&P500 & 1.11 & 4.99 & 69.27 & 57.46 & 79.01 \\ 
  TGARCH-SSTD & S\&P500 & 0.72 & 92.22 & 28.52 & 71.87 & 52.94 \\ 
  GARCH-SSTD & S\&P500 & 1.27 & 20.76 & 35.24 & 43.34 & 28.13 \\ 
  TGARCH-STD & S\&P500 & 1.43 & 0.20 & 14.91 & 46.97 & 27.20 \\ 
  SVR-GARCH-KDE & S\&P500 & 0.79 & 29.34 & 44.83 & 18.28 & 13.71 \\ 
  GARCH-STD & S\&P500 & 1.67 & 16.17 & 2.95 & 65.77 & 6.15 \\ 
  EGARCH-STD & S\&P500 & 1.75 & 5.79 & 1.58 & 37.62 & 3.67 \\ 
  TGARCH-NORM & S\&P500 & 2.15 & 3.39 & 0.04 & 27.64 & 0.10 \\ 
  EGARCH-NORM & S\&P500 & 2.23 & 2.40 & 0.02 & 25.88 & 0.04 \\ 
  GARCH-NORM & S\&P500 & 2.38 & 0.20 & 0.00 & 94.85 & 0.01 \\ 
   \hline
\hline
TGARCH-SSTD & Nikkei225 & 1.21 & 4.99 & 47.61 & 54.48 & 64.58 \\ 
  GARCH-SSTD & Nikkei225 & 1.21 & 2.79 & 47.61 & 54.48 & 64.58 \\ 
  EGARCH-SSTD & Nikkei225 & 1.37 & 4.19 & 21.61 & 49.21 & 36.76 \\ 
  GARCH-STD & Nikkei225 & 1.69 & 1.80 & 2.60 & 66.33 & 5.56 \\ 
  TGARCH-STD & Nikkei225 & 1.77 & 1.00 & 1.37 & 37.30 & 3.23 \\ 
  SVR-GARCH-KDE & Nikkei225 & 0.24 & 59.48 & 0.13 & 90.41 & 0.55 \\ 
  EGARCH-STD & Nikkei225 & 2.09 & 0.80 & 0.07 & 29.17 & 0.19 \\ 
  GARCH-NORM & Nikkei225 & 2.17 & 0.40 & 0.03 & 88.16 & 0.14 \\ 
  TGARCH-NORM & Nikkei225 & 2.33 & 0.20 & 0.01 & 93.19 & 0.03 \\ 
  EGARCH-NORM & Nikkei225 & 2.42 & 0.20 & 0.00 & 95.21 & 0.01 \\ 
   \hline
\hline
SVR-GARCH-KDE & EuroStoxx50 & 1.53 & 76.25 & 8.17 & 44.23 & 16.36 \\ 
  GARCH-SSTD & EuroStoxx50 & 1.69 & 72.85 & 2.60 & 39.54 & 5.84 \\ 
  TGARCH-SSTD & EuroStoxx50 & 1.69 & 43.51 & 2.60 & 66.33 & 5.56 \\ 
  TGARCH-STD & EuroStoxx50 & 1.77 & 1.60 & 1.37 & 70.53 & 3.39 \\ 
  GARCH-STD & EuroStoxx50 & 1.85 & 1.80 & 0.70 & 35.15 & 1.70 \\ 
  TGARCH-NORM & EuroStoxx50 & 2.09 & 0.20 & 0.07 & 85.17 & 0.28 \\ 
  GARCH-NORM & EuroStoxx50 & 2.17 & 6.19 & 0.03 & 27.33 & 0.08 \\ 
  EGARCH-STD & EuroStoxx50 & 2.33 & 9.38 & 0.01 & 93.19 & 0.03 \\ 
  EGARCH-SSTD & EuroStoxx50 & 2.42 & 4.79 & 0.00 & 95.21 & 0.01 \\ 
  EGARCH-NORM & EuroStoxx50 & 2.50 & 3.19 & 0.00 & 96.87 & 0.00 \\ 
   \hline
\hline
\end{tabular}
\endgroup
\caption{Results for one-day-ahead VaR forecasts from July 1, 2011 to June 30, 2016 for $\alpha =  0.01$. UC, ID, CC and SPA indicate the p-value of the corresponding test. The results are in descending order with respect to CC for each index. All values in the columns Quantile, Violations, UC, ID, CC and SPA are given in percent.} 
\label{Tab:case1_1}
\end{table}

\begin{table}[ht]
\centering
\begingroup\fontsize{11pt}{12pt}\selectfont
\begin{tabular}{l|c|c|r|r|r|r}
  \hline
\textbf{Model} & \textbf{Index} & \textbf{Violations} & \textbf{SPA} & \textbf{UC} & \textbf{ID} & \textbf{CC} \\ 
  \hline
\hline
GARCH-SSTD & S\&P500 & 2.62 & 34.33 & 78.12 & 98.97 & 95.22 \\ 
  EGARCH-SSTD & S\&P500 & 3.02 & 1.40 & 25.18 & 98.94 & 51.31 \\ 
  TGARCH-SSTD & S\&P500 & 2.38 & 80.44 & 79.19 & 22.60 & 46.40 \\ 
  GARCH-STD & S\&P500 & 3.50 & 0.80 & 3.24 & 93.45 & 9.47 \\ 
  GARCH-NORM & S\&P500 & 3.82 & 1.20 & 0.55 & 99.20 & 2.10 \\ 
  SVR-GARCH-KDE & S\&P500 & 1.43 & 34.53 & 0.83 & 52.45 & 1.62 \\ 
  TGARCH-STD & S\&P500 & 3.90 & 0.00 & 0.33 & 75.52 & 1.02 \\ 
  TGARCH-NORM & S\&P500 & 4.05 & 1.60 & 0.12 & 69.37 & 0.36 \\ 
  EGARCH-STD & S\&P500 & 4.37 & 0.00 & 0.01 & 96.11 & 0.06 \\ 
  EGARCH-NORM & S\&P500 & 4.53 & 0.20 & 0.00 & 92.48 & 0.02 \\ 
   \hline
\hline
GARCH-SSTD & Nikkei225 & 2.90 & 43.91 & 38.01 & 99.90 & 67.96 \\ 
  TGARCH-SSTD & Nikkei225 & 2.42 & 71.66 & 84.78 & 22.29 & 46.72 \\ 
  TGARCH-STD & Nikkei225 & 3.14 & 0.40 & 16.44 & 97.67 & 37.15 \\ 
  TGARCH-NORM & Nikkei225 & 3.22 & 1.60 & 11.91 & 96.33 & 28.60 \\ 
  EGARCH-NORM & Nikkei225 & 3.22 & 0.40 & 11.91 & 96.33 & 28.60 \\ 
  GARCH-NORM & Nikkei225 & 3.38 & 0.00 & 5.87 & 92.85 & 15.55 \\ 
  GARCH-STD & Nikkei225 & 3.38 & 0.00 & 5.87 & 89.33 & 14.96 \\ 
  EGARCH-SSTD & Nikkei225 & 3.06 & 13.97 & 22.21 & 12.14 & 14.30 \\ 
  SVR-GARCH-KDE & Nikkei225 & 1.69 & 73.65 & 5.26 & 14.27 & 2.18 \\ 
  EGARCH-STD & Nikkei225 & 3.86 & 0.20 & 0.43 & 99.42 & 1.68 \\ 
   \hline
\hline
TGARCH-SSTD & EuroStoxx50 & 2.98 & 79.84 & 29.36 & 72.90 & 42.00 \\ 
  TGARCH-NORM & EuroStoxx50 & 3.22 & 0.00 & 11.91 & 83.46 & 24.78 \\ 
  TGARCH-STD & EuroStoxx50 & 3.38 & 0.00 & 5.87 & 89.33 & 14.96 \\ 
  GARCH-SSTD & EuroStoxx50 & 3.46 & 68.86 & 3.99 & 91.84 & 11.13 \\ 
  SVR-GARCH-KDE & EuroStoxx50 & 3.14 & 29.74 & 16.44 & 11.43 & 4.35 \\ 
  GARCH-STD & EuroStoxx50 & 3.70 & 5.19 & 1.11 & 97.42 & 3.88 \\ 
  GARCH-NORM & EuroStoxx50 & 3.95 & 5.59 & 0.26 & 75.92 & 0.81 \\ 
  EGARCH-STD & EuroStoxx50 & 4.27 & 1.20 & 0.03 & 88.73 & 0.12 \\ 
  EGARCH-SSTD & EuroStoxx50 & 4.51 & 1.40 & 0.00 & 95.50 & 0.02 \\ 
  EGARCH-NORM & EuroStoxx50 & 4.59 & 1.60 & 0.00 & 97.11 & 0.01 \\ 
   \hline
\hline
\end{tabular}
\endgroup
\caption{Results for one-day-ahead VaR forecasts from July 1, 2011 to June 30, 2016 for $\alpha =  0.025$. UC, ID, CC and SPA indicate the p-value of the corresponding test. The results are in descending order with respect to CC for each index. All values in the columns Quantile, Violations, UC, ID, CC and SPA are given in percent.} 
\label{Tab:case1_2.5}
\end{table}

\begin{table}[ht]
\centering
\begingroup\fontsize{11pt}{12pt}\selectfont
\begin{tabular}{l|c|c|r|r|r|r}
  \hline
\textbf{Model} & \textbf{Index} & \textbf{Violations} & \textbf{SPA} & \textbf{UC} & \textbf{ID} & \textbf{CC} \\ 
  \hline
\hline
GARCH-SSTD & S\&P500 & 5.01 & 57.49 & 98.97 & 89.07 & 89.06 \\ 
  GARCH-NORM & S\&P500 & 5.72 & 0.00 & 24.94 & 99.79 & 51.40 \\ 
  TGARCH-SSTD & S\&P500 & 5.41 & 87.82 & 51.47 & 60.38 & 48.83 \\ 
  GARCH-STD & S\&P500 & 5.80 & 2.20 & 20.21 & 99.23 & 43.99 \\ 
  SVR-GARCH-KDE & S\&P500 & 4.37 & 24.95 & 29.68 & 30.25 & 17.55 \\ 
  EGARCH-SSTD & S\&P500 & 6.44 & 1.20 & 2.47 & 83.92 & 6.73 \\ 
  TGARCH-NORM & S\&P500 & 6.60 & 0.40 & 1.30 & 77.88 & 3.56 \\ 
  EGARCH-NORM & S\&P500 & 6.76 & 1.00 & 0.65 & 94.35 & 2.33 \\ 
  TGARCH-STD & S\&P500 & 6.76 & 0.00 & 0.65 & 71.36 & 1.77 \\ 
  EGARCH-STD & S\&P500 & 7.00 & 0.80 & 0.21 & 87.46 & 0.78 \\ 
   \hline
\hline
EGARCH-SSTD & Nikkei225 & 5.23 & 15.17 & 70.78 & 97.26 & 90.66 \\ 
  SVR-GARCH-KDE & Nikkei225 & 4.59 & 27.74 & 50.10 & 97.11 & 77.43 \\ 
  TGARCH-SSTD & Nikkei225 & 4.67 & 94.81 & 58.95 & 89.41 & 77.30 \\ 
  GARCH-SSTD & Nikkei225 & 5.07 & 34.93 & 90.69 & 75.29 & 74.78 \\ 
  TGARCH-NORM & Nikkei225 & 5.23 & 0.20 & 70.78 & 68.85 & 64.18 \\ 
  TGARCH-STD & Nikkei225 & 5.56 & 0.20 & 37.71 & 89.62 & 60.67 \\ 
  GARCH-NORM & Nikkei225 & 5.88 & 4.99 & 16.68 & 78.41 & 30.16 \\ 
  EGARCH-NORM & Nikkei225 & 5.72 & 0.60 & 25.68 & 49.18 & 25.85 \\ 
  GARCH-STD & Nikkei225 & 6.04 & 0.00 & 10.34 & 71.98 & 19.11 \\ 
  EGARCH-STD & Nikkei225 & 6.04 & 0.60 & 10.34 & 71.98 & 19.11 \\ 
   \hline
\hline
TGARCH-SSTD & EuroStoxx50 & 5.15 & 89.62 & 80.55 & 65.27 & 63.32 \\ 
  GARCH-SSTD & EuroStoxx50 & 5.96 & 57.29 & 13.21 & 95.83 & 30.84 \\ 
  GARCH-NORM & EuroStoxx50 & 5.96 & 48.90 & 13.21 & 95.83 & 30.84 \\ 
  TGARCH-NORM & EuroStoxx50 & 6.04 & 2.59 & 10.34 & 78.10 & 20.73 \\ 
  SVR-GARCH-KDE & EuroStoxx50 & 5.48 & 33.53 & 44.90 & 26.86 & 20.17 \\ 
  EGARCH-SSTD & EuroStoxx50 & 6.12 & 1.40 & 7.99 & 68.65 & 14.82 \\ 
  TGARCH-STD & EuroStoxx50 & 6.20 & 0.00 & 6.11 & 59.29 & 10.26 \\ 
  EGARCH-NORM & EuroStoxx50 & 6.52 & 1.80 & 1.85 & 99.10 & 6.18 \\ 
  GARCH-STD & EuroStoxx50 & 6.60 & 3.99 & 1.33 & 96.58 & 4.52 \\ 
  EGARCH-STD & EuroStoxx50 & 6.84 & 0.80 & 0.46 & 69.51 & 1.27 \\ 
   \hline
\hline
\end{tabular}
\endgroup
\caption{Results for one-day-ahead VaR forecasts from July 1, 2011 to June 30, 2016 for $\alpha =  0.05$. UC, ID, CC and SPA indicate the p-value of the corresponding test. The results are in descending order with respect to CC for each index. All values in the columns Quantile, Violations, UC, ID, CC and SPA are given in percent.} 
\label{Tab:case1_5}
\end{table}

\begin{table}[ht]
\centering
\begingroup\fontsize{11pt}{12pt}\selectfont
\begin{tabular}{l|c|c|r|r|r|r}
  \hline
\textbf{Model} & \textbf{Index} & \textbf{Violations} & \textbf{SPA} & \textbf{UC} & \textbf{ID} & \textbf{CC} \\ 
  \hline
\hline
TGARCH-SSTD & S\&P500 & 1.03 & 11.78 & 90.58 & 30.09 & 29.88 \\ 
  SVR-GARCH-KDE & S\&P500 & 0.95 & 94.01 & 86.85 & 25.94 & 25.59 \\ 
  GARCH-SSTD & S\&P500 & 1.19 & 10.38 & 50.57 & 3.91 & 3.14 \\ 
  TGARCH-STD & S\&P500 & 1.99 & 1.80 & 0.19 & 25.26 & 0.21 \\ 
  GARCH-STD & S\&P500 & 1.99 & 2.99 & 0.19 & 25.26 & 0.21 \\ 
  TGARCH-NORM & S\&P500 & 2.31 & 3.59 & 0.01 & 39.72 & 0.01 \\ 
  GARCH-NORM & S\&P500 & 2.54 & 3.39 & 0.00 & 51.83 & 0.00 \\ 
  EGARCH-STD & S\&P500 & 2.78 & 0.00 & 0.00 & 64.14 & 0.00 \\ 
  EGARCH-SSTD & S\&P500 & 2.70 & 0.80 & 0.00 & 4.50 & 0.00 \\ 
  EGARCH-NORM & S\&P500 & 3.26 & 0.60 & 0.00 & 14.87 & 0.00 \\ 
   \hline
\hline
TGARCH-SSTD & Nikkei225 & 1.45 & 3.19 & 13.59 & 52.97 & 17.42 \\ 
  GARCH-SSTD & Nikkei225 & 1.45 & 2.40 & 13.59 & 52.97 & 17.42 \\ 
  SVR-GARCH-KDE & Nikkei225 & 0.32 & 57.29 & 0.51 & 87.23 & 1.97 \\ 
  GARCH-STD & Nikkei225 & 1.85 & 0.80 & 0.70 & 19.57 & 0.51 \\ 
  TGARCH-STD & Nikkei225 & 1.93 & 1.80 & 0.34 & 22.56 & 0.31 \\ 
  GARCH-NORM & Nikkei225 & 2.01 & 1.80 & 0.16 & 25.77 & 0.18 \\ 
  TGARCH-NORM & Nikkei225 & 2.09 & 2.20 & 0.07 & 5.45 & 0.02 \\ 
  EGARCH-STD & Nikkei225 & 3.95 & 0.20 & 0.00 & 75.92 & 0.00 \\ 
  EGARCH-SSTD & Nikkei225 & 2.50 & 1.20 & 0.00 & 48.53 & 0.00 \\ 
  EGARCH-NORM & Nikkei225 & 2.58 & 0.80 & 0.00 & 15.65 & 0.00 \\ 
   \hline
\hline
SVR-GARCH-KDE & EuroStoxx50 & 0.97 & 64.07 & 90.41 & 62.85 & 88.31 \\ 
  GARCH-SSTD & EuroStoxx50 & 2.01 & 8.78 & 0.16 & 25.77 & 0.18 \\ 
  TGARCH-SSTD & EuroStoxx50 & 2.33 & 5.99 & 0.01 & 40.43 & 0.01 \\ 
  TGARCH-STD & EuroStoxx50 & 2.58 & 4.99 & 0.00 & 15.65 & 0.00 \\ 
  TGARCH-NORM & EuroStoxx50 & 2.66 & 4.79 & 0.00 & 18.07 & 0.00 \\ 
  GARCH-STD & EuroStoxx50 & 2.33 & 5.79 & 0.01 & 9.64 & 0.00 \\ 
  GARCH-NORM & EuroStoxx50 & 2.58 & 0.80 & 0.00 & 15.65 & 0.00 \\ 
  EGARCH-STD & EuroStoxx50 & 4.03 & 1.40 & 0.00 & 17.32 & 0.00 \\ 
  EGARCH-SSTD & EuroStoxx50 & 3.70 & 1.80 & 0.00 & 2.49 & 0.00 \\ 
  EGARCH-NORM & EuroStoxx50 & 3.30 & 0.20 & 0.00 & 44.47 & 0.00 \\ 
   \hline
\hline
\end{tabular}
\endgroup
\caption{Results for ten-days-ahead VaR forecasts from July 1, 2011 to June 30, 2016 for $\alpha =  0.01$. UC, ID, CC and SPA indicate the p-value of the corresponding test. The results are in descending order with respect to CC for each index. All values in the columns Quantile, Violations, UC, ID, CC and SPA are given in percent.} 
\label{Tab:case1_10days_1}
\end{table}

\begin{table}[ht]
\centering
\begingroup\fontsize{11pt}{12pt}\selectfont
\begin{tabular}{l|c|c|r|r|r|r}
  \hline
\textbf{Model} & \textbf{Index} & \textbf{Violations} & \textbf{SPA} & \textbf{UC} & \textbf{ID} & \textbf{CC} \\ 
  \hline
\hline
TGARCH-SSTD & S\&P500 & 2.62 & 13.57 & 78.12 & 55.96 & 53.84 \\ 
  GARCH-SSTD & S\&P500 & 2.78 & 6.99 & 52.88 & 22.94 & 18.81 \\ 
  GARCH-STD & S\&P500 & 3.18 & 3.39 & 13.82 & 39.68 & 13.22 \\ 
  TGARCH-STD & S\&P500 & 3.42 & 0.20 & 4.80 & 51.32 & 7.27 \\ 
  GARCH-NORM & S\&P500 & 3.58 & 2.79 & 2.14 & 59.33 & 4.20 \\ 
  SVR-GARCH-KDE & S\&P500 & 1.75 & 95.21 & 7.15 & 16.45 & 3.24 \\ 
  TGARCH-NORM & S\&P500 & 3.58 & 0.40 & 2.14 & 25.03 & 1.77 \\ 
  EGARCH-STD & S\&P500 & 4.37 & 0.00 & 0.01 & 30.25 & 0.02 \\ 
  EGARCH-SSTD & S\&P500 & 4.77 & 0.20 & 0.00 & 22.74 & 0.00 \\ 
  EGARCH-NORM & S\&P500 & 4.61 & 0.20 & 0.00 & 17.75 & 0.00 \\ 
   \hline
\hline
GARCH-STD & Nikkei225 & 2.74 & 0.20 & 59.74 & 60.95 & 53.02 \\ 
  GARCH-SSTD & Nikkei225 & 2.25 & 8.58 & 57.31 & 36.53 & 31.17 \\ 
  GARCH-NORM & Nikkei225 & 3.30 & 0.00 & 8.45 & 86.53 & 19.54 \\ 
  TGARCH-SSTD & Nikkei225 & 2.74 & 0.00 & 59.74 & 20.72 & 18.02 \\ 
  TGARCH-STD & Nikkei225 & 3.06 & 0.00 & 22.21 & 33.30 & 15.80 \\ 
  TGARCH-NORM & Nikkei225 & 3.46 & 0.00 & 3.99 & 52.40 & 6.35 \\ 
  EGARCH-SSTD & Nikkei225 & 4.03 & 0.40 & 0.15 & 43.03 & 0.28 \\ 
  EGARCH-NORM & Nikkei225 & 4.11 & 0.00 & 0.09 & 46.87 & 0.19 \\ 
  SVR-GARCH-KDE & Nikkei225 & 1.29 & 48.10 & 0.26 & 5.17 & 0.06 \\ 
  EGARCH-STD & Nikkei225 & 5.31 & 0.20 & 0.00 & 96.40 & 0.00 \\ 
   \hline
\hline
TGARCH-SSTD & EuroStoxx50 & 3.22 & 50.30 & 11.91 & 40.62 & 12.06 \\ 
  GARCH-SSTD & EuroStoxx50 & 3.30 & 50.10 & 8.45 & 44.47 & 10.04 \\ 
  TGARCH-NORM & EuroStoxx50 & 3.38 & 9.38 & 5.87 & 48.40 & 8.11 \\ 
  GARCH-STD & EuroStoxx50 & 3.46 & 8.78 & 3.99 & 52.40 & 6.35 \\ 
  SVR-GARCH-KDE & EuroStoxx50 & 3.06 & 85.03 & 22.21 & 9.73 & 4.61 \\ 
  TGARCH-STD & EuroStoxx50 & 3.46 & 13.77 & 3.99 & 20.23 & 2.45 \\ 
  GARCH-NORM & EuroStoxx50 & 3.78 & 0.20 & 0.70 & 68.40 & 1.80 \\ 
  EGARCH-STD & EuroStoxx50 & 5.88 & 0.40 & 0.00 & 43.78 & 0.00 \\ 
  EGARCH-SSTD & EuroStoxx50 & 4.91 & 0.20 & 0.00 & 3.67 & 0.00 \\ 
  EGARCH-NORM & EuroStoxx50 & 4.83 & 0.00 & 0.00 & 23.77 & 0.00 \\ 
   \hline
\hline
\end{tabular}
\endgroup
\caption{Results for ten-days-ahead VaR forecasts from July 1, 2011 to June 30, 2016 for $\alpha =  0.025$. UC, ID, CC and SPA indicate the p-value of the corresponding test. The results are in descending order with respect to CC for each index. All values in the columns Quantile, Violations, UC, ID, CC and SPA are given in percent.} 
\label{Tab:case1_10days_2.5}
\end{table}

\begin{table}[ht]
\centering
\begingroup\fontsize{11pt}{12pt}\selectfont
\begin{tabular}{l|c|c|r|r|r|r}
  \hline
\textbf{Model} & \textbf{Index} & \textbf{Violations} & \textbf{SPA} & \textbf{UC} & \textbf{ID} & \textbf{CC} \\ 
  \hline
\hline
TGARCH-NORM & S\&P500 & 5.09 & 1.20 & 88.72 & 34.98 & 34.63 \\ 
  TGARCH-STD & S\&P500 & 5.01 & 0.00 & 98.97 & 31.66 & 31.66 \\ 
  GARCH-SSTD & S\&P500 & 4.29 & 2.20 & 23.85 & 57.38 & 28.64 \\ 
  GARCH-STD & S\&P500 & 5.17 & 0.00 & 78.70 & 17.88 & 17.24 \\ 
  TGARCH-SSTD & S\&P500 & 4.21 & 4.39 & 18.86 & 24.22 & 10.20 \\ 
  GARCH-NORM & S\&P500 & 5.48 & 0.00 & 43.69 & 4.89 & 3.62 \\ 
  EGARCH-NORM & S\&P500 & 6.52 & 0.00 & 1.80 & 29.69 & 1.81 \\ 
  EGARCH-STD & S\&P500 & 6.60 & 0.00 & 1.30 & 32.87 & 1.50 \\ 
  SVR-GARCH-KDE & S\&P500 & 3.34 & 56.89 & 0.41 & 17.11 & 0.28 \\ 
  EGARCH-SSTD & S\&P500 & 7.23 & 0.00 & 0.06 & 62.62 & 0.18 \\ 
   \hline
\hline
TGARCH-SSTD & Nikkei225 & 5.23 & 10.18 & 70.78 & 94.63 & 88.21 \\ 
  GARCH-SSTD & Nikkei225 & 4.99 & 21.16 & 98.96 & 87.41 & 87.40 \\ 
  TGARCH-STD & Nikkei225 & 5.31 & 8.18 & 61.51 & 72.91 & 64.25 \\ 
  TGARCH-NORM & Nikkei225 & 5.31 & 12.18 & 61.51 & 72.91 & 64.25 \\ 
  GARCH-NORM & Nikkei225 & 5.72 & 5.99 & 25.68 & 99.95 & 52.55 \\ 
  GARCH-STD & Nikkei225 & 5.72 & 8.18 & 25.68 & 89.19 & 46.89 \\ 
  SVR-GARCH-KDE & Nikkei225 & 3.95 & 88.82 & 7.70 & 75.92 & 15.90 \\ 
  EGARCH-SSTD & Nikkei225 & 6.36 & 0.60 & 3.44 & 90.35 & 9.65 \\ 
  EGARCH-NORM & Nikkei225 & 6.12 & 3.59 & 7.99 & 31.60 & 6.82 \\ 
  EGARCH-STD & Nikkei225 & 7.97 & 0.00 & 0.00 & 91.62 & 0.00 \\ 
   \hline
\hline
SVR-GARCH-KDE & EuroStoxx50 & 5.07 & 50.10 & 90.69 & 61.34 & 60.92 \\ 
  TGARCH-NORM & EuroStoxx50 & 6.12 & 2.59 & 7.99 & 6.55 & 1.41 \\ 
  TGARCH-SSTD & EuroStoxx50 & 6.12 & 4.99 & 7.99 & 2.43 & 0.53 \\ 
  TGARCH-STD & EuroStoxx50 & 6.36 & 0.00 & 3.44 & 4.28 & 0.46 \\ 
  GARCH-NORM & EuroStoxx50 & 6.36 & 2.20 & 3.44 & 4.28 & 0.46 \\ 
  GARCH-SSTD & EuroStoxx50 & 6.52 & 2.99 & 1.85 & 6.06 & 0.38 \\ 
  GARCH-STD & EuroStoxx50 & 6.76 & 0.40 & 0.67 & 9.76 & 0.25 \\ 
  EGARCH-STD & EuroStoxx50 & 9.10 & 0.20 & 0.00 & 1.03 & 0.00 \\ 
  EGARCH-SSTD & EuroStoxx50 & 7.41 & 0.20 & 0.03 & 0.13 & 0.00 \\ 
  EGARCH-NORM & EuroStoxx50 & 7.17 & 0.60 & 0.10 & 0.18 & 0.00 \\ 
   \hline
\hline
\end{tabular}
\endgroup
\caption{Results for ten-days-ahead VaR forecasts from July 1, 2011 to June 30, 2016 for $\alpha =  0.05$. UC, ID, CC and SPA indicate the p-value of the corresponding test. The results are in descending order with respect to CC for each index. All values in the columns Quantile, Violations, UC, ID, CC and SPA are given in percent.} 
\label{Tab:case1_10days_5}
\end{table}


\clearpage

\section{Conclusion}\label{Sec:Conc}

In a large-scale empirical comparison \cite{Kuester.2005} find VaR models belonging to the location-scale class superior to alternative approaches. However, the location-scale models considered in their study are parametric and based on distributional assumptions. Motivated by the potential shortcomings of a parametric approach, the paper introduces a nonparametric and nonlinear VaR forecasting framework based on the location-scale class. The mean and volatility model are modeled with SVR in an ARMA and GARCH like fashion, respectively. In addition, the VaR forecast is obtained by estimating the distribution function of the standardized residuals via KDE. 

To evaluate the performance of the SVR-GARCH-KDE hybrid, VaR is forecasted for three indices: Euro STOXX 50, Nikkei 225 and S\&P 500, considering the different quantiles of $\alpha \in \{0.01, 0.025, 0.05\}$ for forecast horizons of one and ten days. GARCH, EGARCH and TGARCH models coupled with the normal, $t$- and skewed $t$-distribution serve as benchmarks and are compared to the proposed model using a LR testing framework for interval forecasts \citep{Christoffersen.1998} and the SPA test of \cite{Hansen.2005}. Grid search with the goal to maximize the $p$-value of the LR test for conditional coverage is used to set the SVR parameters. The SVR-GARCH-KDE hybrid delivers competitive results. For instance, in case of the one-day-ahead forecast horizon it is the best model for the Euro STOXX 50 for $\alpha= 0.01$ and the third best model overall. The TGARCH and GARCH model in combination with the skewed $t$-distribution show the best results. In contrast to the benchmark models, which usually underestimate risk, the SVR-GARCH-KDE hybrid tends to overestimate risk. This can lead to situations where the SVR-GARCH-KDE hybrid has an average performance regarding the $p$-value of the test for conditional coverage. However, with respect to risk management, the use of the SVR-GARCH-KDE might be still favorable over using the benchmarks since all approaches exhibit statistical uncertainty. The tendency to overestimate risk can, therefore, serve as a model risk buffer. This is supported by the results of the SPA test that evaluates the loss function which penalizes especially large VaR violations. For instance, the proposed model has the best performance for the ten-days-ahead forecast horizon. 

In general, the competitive results indicate that the proposed SVR-GARCH-KDE hybrid is a promising alternative. Despite fixing and not retuning the hyperparameters for five years, it is among the top three models with respect to conditional coverage and loss minimization measured by the SPA test for a forecast horizon of one as well as ten days. Additionally, it is the best model in minimizing the loss function for ten-days-ahead forecasts. This indicates the proposed SVR-GARCH-KDE hybrid is a robust VaR modeling approach that is capable of capturing complex nonlinear structures in the volatility process and has the flexibility to model a wide class of tail events. Moreover, further improvements can be expected by refining the tuning routine. However, there exist several ways that can lead to an improved performance. First, tuning could be done for more parameters. For instance, in the KDE part of the estimation procedure, the kernel function and bandwidth estimator are set without tuning. Hence, considering different kernel functions and more flexible bandwidth estimators are potential ways to improve the performance further. Moreover, the kernel in the SVR part is also fixed and could be varied. Second, more recent information could be used in the parameter selection by re-tuning the model. Here, tuning is done for a block of five years of data. Then, based on the optimal parameters found for this data block, forecasts for five years are made and the parameters are held fixed. Thus, annual or even shorter re-tuning periods could result in parameters that are more appropriate for the existing market risk. Additionally, refining the grid can also result in better parameter choices.

In addition to modifying the tuning routine, the SVR-GARCH-KDE hybrid could be improved by changing the model specification. Overall, the TGARCH model with the skewed $t$-distribution achieves very good results. Hence, the SVR-GARCH-KDE hybrid could be modified such that it also accounts for asymmetric reactions of the volatility to past returns in a TGARCH like manner. Moreover, the proposed procedure does not ensure that the estimated variances are positive. Here, this problem is handled by replacing non-positive estimates with the last positive. However, positive variance estimates could be ensured by modeling the logarithm of the squared mean model residuals instead. 

All above mentioned adjustments are potential starting points for future research to further improve the proposed framework. However, although the suggested modifications of the tuning procedure are reasonable approaches to improve the model performance, they also increase the computational complexity. After all, this is a slight drawback of the SVR-GARCH-KDE hybrid in comparison to standard models. It is, however offset by potential performance gains.

\clearpage

\clearpage

\bibliographystyle{spbasic}
\bibliography{literature_v006}

\end{document}